\documentclass[twocolumn]{aa}
\usepackage{graphicx}
\usepackage{amsmath,amsfonts,amssymb}
\usepackage{txfonts}
\usepackage{color}
\usepackage{natbib}
\usepackage{float}
\usepackage{dblfloatfix}
\usepackage{afterpage}
\usepackage{ifthen}
\usepackage[morefloats=12]{morefloats}
\usepackage{placeins}
\bibpunct{(}{)}{;}{a}{}{,}
\usepackage[switch]{lineno}
\definecolor{linkcolor}{rgb}{0.6,0,0}
\definecolor{citecolor}{rgb}{0,0,0.75}
\definecolor{urlcolor}{rgb}{0.12,0.46,0.7}
\usepackage[breaklinks, colorlinks, urlcolor=urlcolor,
    linkcolor=linkcolor,citecolor=citecolor,pdfencoding=auto]{hyperref}
\hypersetup{linktocpage}
\usepackage{tabularx}
\usepackage[ruled,vlined]{algorithm2e}
\usepackage[percent]{overpic}

\def\setsymbol#1#2{\expandafter\def\csname #1\endcsname{#2}}
\def\getsymbol#1{\csname #1\endcsname}

\def\Planck{\textit{Planck}}

\newbox\tablebox    \newdimen\tablewidth
\def\leaderfil{\leaders\hbox to 5pt{\hss.\hss}\hfil}
\def\endPlancktablewide{\tablewidth=\textwidth 
    $$\hss\copy\tablebox\hss$$
    \vskip-\lastskip\vskip -2pt}
\def\tablenote#1 #2\par{\begingroup \parindent=0.8em
    \abovedisplayshortskip=0pt\belowdisplayshortskip=0pt
    \noindent
    $$\hss\vbox{\hsize\tablewidth \hangindent=\parindent \hangafter=1 \noindent
    \hbox to \parindent{$^#1$\hss}\strut#2\strut\par}\hss$$
    \endgroup}
\def\doubleline{\vskip 3pt\hrule \vskip 1.5pt \hrule \vskip 5pt}

\def\L2{\ifmmode L_2\else $L_2$\fi}

\def\DeltaT{\ifmmode \Delta T\else $\Delta T$\fi}
\def\deltat{\ifmmode \Delta t\else $\Delta t$\fi}
\def\fknee{\ifmmode f_{\rm knee}\else $f_{\rm knee}$\fi}
\def\Fmax{\ifmmode F_{\rm max}\else $F_{\rm max}$\fi}
\def\solar{\ifmmode{\rm M}_{\mathord\odot}\else${\rm M}_{\mathord\odot}$\fi}
\def\Msolar{\ifmmode{\rm M}_{\mathord\odot}\else${\rm M}_{\mathord\odot}$\fi}
\def\Lsolar{\ifmmode{\rm L}_{\mathord\odot}\else${\rm L}_{\mathord\odot}$\fi}
\def\inv{\ifmmode^{-1}\else$^{-1}$\fi}
\def\mo{\ifmmode^{-1}\else$^{-1}$\fi}
\def\sup#1{\ifmmode ^{\rm #1}\else $^{\rm #1}$\fi}
\def\expo#1{\ifmmode \times 10^{#1}\else $\times 10^{#1}$\fi}
\def\,{\thinspace}
\def\lsim{\mathrel{\raise .4ex\hbox{\rlap{$<$}\lower 1.2ex\hbox{$\sim$}}}}
\def\gsim{\mathrel{\raise .4ex\hbox{\rlap{$>$}\lower 1.2ex\hbox{$\sim$}}}}

\def\simprop{\mathrel{\raise .4ex\hbox{\rlap{$\propto$}\lower 1.2ex\hbox{$\sim$}}}}
\def\deg{\ifmmode^\circ\else$^\circ$\fi}
\def\pdeg{\ifmmode $\setbox0=\hbox{$^{\circ}$}\rlap{\hskip.11\wd0 .}$^{\circ}
          \else \setbox0=\hbox{$^{\circ}$}\rlap{\hskip.11\wd0 .}$^{\circ}$\fi}
\def\arcs{\ifmmode {^{\scriptstyle\prime\prime}}
          \else $^{\scriptstyle\prime\prime}$\fi}
\def\arcm{\ifmmode {^{\scriptstyle\prime}}
          \else $^{\scriptstyle\prime}$\fi}
\newdimen\sa  \newdimen\sb
\def\parcs{\sa=.07em \sb=.03em
     \ifmmode \hbox{\rlap{.}}^{\scriptstyle\prime\kern -\sb\prime}\hbox{\kern -\sa}
     \else \rlap{.}$^{\scriptstyle\prime\kern -\sb\prime}$\kern -\sa\fi}
\def\parcm{\sa=.08em \sb=.03em
     \ifmmode \hbox{\rlap{.}\kern\sa}^{\scriptstyle\prime}\hbox{\kern-\sb}
     \else \rlap{.}\kern\sa$^{\scriptstyle\prime}$\kern-\sb\fi}
\def\ra[#1 #2 #3.#4]{#1\sup{h}#2\sup{m}#3\sup{s}\llap.#4}
\def\dec[#1 #2 #3.#4]{#1\deg#2\arcm#3\arcs\llap.#4}
\def\deco[#1 #2 #3]{#1\deg#2\arcm#3\arcs}
\def\rra[#1 #2]{#1\sup{h}#2\sup{m}}

\def\dots{\relax\ifmmode \ldots\else $\ldots$\fi}
\def\WHzsr{\ifmmode $W\,Hz\mo\,sr\mo$\else W\,Hz\mo\,sr\mo\fi}
\def\mHz{\ifmmode $\,mHz$\else \,mHz\fi}
\def\GHz{\ifmmode $\,GHz$\else \,GHz\fi}
\def\mKs{\ifmmode $\,mK\,s$^{1/2}\else \,mK\,s$^{1/2}$\fi}
\def\muKs{\ifmmode \,\mu$K\,s$^{1/2}\else \,$\mu$K\,s$^{1/2}$\fi}
\def\muKRJs{\ifmmode \,\mu$K$_{\rm RJ}$\,s$^{1/2}\else \,$\mu$K$_{\rm RJ}$\,s$^{1/2}$\fi}
\def\muKHz{\ifmmode \,\mu$K\,Hz$^{-1/2}\else \,$\mu$K\,Hz$^{-1/2}$\fi}
\def\MJysr{\ifmmode \,$MJy\,sr\mo$\else \,MJy\,sr\mo\fi}
\def\MJysrmK{\ifmmode \,$MJy\,sr\mo$\,mK$_{\rm CMB}\mo\else \,MJy\,sr\mo\,mK$_{\rm CMB}\mo$\fi}
\def\microns{\ifmmode \,\mu$m$\else \,$\mu$m\fi}

\def\muK{\ifmmode \,\mu$K$\else \,$\mu$\hbox{K}\fi}
\def\microK{\ifmmode \,\mu$K$\else \,$\mu$\hbox{K}\fi}
\def\muW{\ifmmode \,\mu$W$\else \,$\mu$\hbox{W}\fi}
\def\kms{\ifmmode $\,km\,s$^{-1}\else \,km\,s$^{-1}$\fi}
\def\kmsMpc{\ifmmode $\,\kms\,Mpc\mo$\else \,\kms\,Mpc\mo\fi}
\providecommand{\sorthelp}[1]{}

\def\WMAP{\textit{WMAP}}
\def\COBE{\textit{COBE}}

\def\nside{N_{\mathrm{side}}}
\def\chisq{$\chi^2$}

\def\healpix{\texttt{HEALPix}}
\def\commander{\texttt{Commander}}

\def\smica{\texttt{SMICA}}

\renewcommand{\d}[0]{\vec{d}}

\newcommand{\A}[0]{\tens{A}}

\newcommand{\n}[0]{\vec{n}}

\definecolor{orange}{RGB}{255,127,0}

\newcommand{\s}[0]{\vec{s}}
\renewcommand{\a}[0]{\vec{a}}
\newcommand{\m}[0]{\vec{m}}

\newcommand{\B}[0]{\tens{B}}

\renewcommand{\L}[0]{\tens{L}}
\newcommand{\g}[0]{\vec{g}}

\newcommand{\N}[0]{\tens{N}}

\renewcommand{\C}[0]{\tens{C}}

\newcommand{\M}[0]{\tens{M}}

\renewcommand{\S}[0]{\tens{S}}
\renewcommand{\r}[0]{\vec{r}}

\renewcommand{\P}[0]{\tens{P}}

\newcommand{\Dbp}[0]{\Delta_{\mathrm{bp}}}

\newcommand{\BP}{\textsc{BeyondPlanck}}

\def\inv{^{-1}}

\begin{document}
 
\title{\bfseries{\scshape{BeyondPlanck}} XV. Polarized foreground
  emission\\ between 30 and 70\,GHz}

\newcommand{\nersc}[0]{1}
\newcommand{\princeton}[0]{2}
\newcommand{\helsinkiA}[0]{3}
\newcommand{\milanoA}[0]{4}
\newcommand{\triesteA}[0]{5}
\newcommand{\haverford}[0]{6}
\newcommand{\helsinkiB}[0]{7}
\newcommand{\triesteB}[0]{8}
\newcommand{\milanoB}[0]{9}
\newcommand{\milanoC}[0]{10}
\newcommand{\oslo}[0]{11}
\newcommand{\jpl}[0]{12}
\newcommand{\mpa}[0]{13}
\newcommand{\planetek}[0]{14}
\author{\small
T.~L.~Svalheim\inst{\oslo}\thanks{Corresponding author: T.~L.~Svalheim; \url{t.l.svalheim@astro.uio.no}}
\and
K.~J.~Andersen\inst{\oslo}
\and
\textcolor{black}{R.~Aurlien}\inst{\oslo}
\and
\textcolor{black}{R.~Banerji}\inst{\oslo}
\and
M.~Bersanelli\inst{\milanoA, \milanoB, \milanoC}
\and
S.~Bertocco\inst{\triesteB}
\and
M.~Brilenkov\inst{\oslo}
\and
M.~Carbone\inst{\planetek}
\and
L.~P.~L.~Colombo\inst{\milanoA}
\and
H.~K.~Eriksen\inst{\oslo}
\and
\textcolor{black}{M.~K.~Foss}\inst{\oslo}
\and
C.~Franceschet\inst{\milanoA, \milanoC}
\and
\textcolor{black}{U.~Fuskeland}\inst{\oslo}
\and
S.~Galeotta\inst{\triesteB}
\and
M.~Galloway\inst{\oslo}
\and
S.~Gerakakis\inst{\planetek}
\and
E.~Gjerl{\o}w\inst{\oslo}
\and
\textcolor{black}{B.~Hensley}\inst{\princeton}
\and
\textcolor{black}{D.~Herman}\inst{\oslo}
\and
M.~Iacobellis\inst{\planetek}
\and
M.~Ieronymaki\inst{\planetek}
\and
\textcolor{black}{H.~T.~Ihle}\inst{\oslo}
\and
J.~B.~Jewell\inst{\jpl}
\and
\textcolor{black}{A.~Karakci}\inst{\oslo}
\and
E.~Keih\"{a}nen\inst{\helsinkiA, \helsinkiB}
\and
R.~Keskitalo\inst{\nersc}
\and
G.~Maggio\inst{\triesteB}
\and
D.~Maino\inst{\milanoA, \milanoB, \milanoC}
\and
M.~Maris\inst{\triesteB}
\and
S.~Paradiso\inst{\milanoA, \milanoC}
\and
B.~Partridge\inst{\haverford}
\and
M.~Reinecke\inst{\mpa}
\and
A.-S.~Suur-Uski\inst{\helsinkiA, \helsinkiB}
\and
D.~Tavagnacco\inst{\triesteB, \triesteA}
\and
H.~Thommesen\inst{\oslo}
\and
D.~J.~Watts\inst{\oslo}
\and
I.~K.~Wehus\inst{\oslo}
\and
A.~Zacchei\inst{\triesteB}
}
\institute{\small
Computational Cosmology Center, Lawrence Berkeley National Laboratory, Berkeley, California, U.S.A.\goodbreak
\and
Department of Astrophysical Sciences, Princeton University, Princeton, NJ 08544,
U.S.A.\goodbreak
\and
Department of Physics, Gustaf H\"{a}llstr\"{o}min katu 2, University of Helsinki, Helsinki, Finland\goodbreak
\and
Dipartimento di Fisica, Universit\`{a} degli Studi di Milano, Via Celoria, 16, Milano, Italy\goodbreak
\and
Dipartimento di Fisica, Universit\`{a} degli Studi di Trieste, via A. Valerio 2, Trieste, Italy\goodbreak
\and
Haverford College Astronomy Department, 370 Lancaster Avenue,
Haverford, Pennsylvania, U.S.A.\goodbreak
\and
Helsinki Institute of Physics, Gustaf H\"{a}llstr\"{o}min katu 2, University of Helsinki, Helsinki, Finland\goodbreak
\and
INAF - Osservatorio Astronomico di Trieste, Via G.B. Tiepolo 11, Trieste, Italy\goodbreak
\and
INAF/IASF Milano, Via E. Bassini 15, Milano, Italy\goodbreak
\and
INFN, Sezione di Milano, Via Celoria 16, Milano, Italy\goodbreak
\and
Institute of Theoretical Astrophysics, University of Oslo, Blindern, Oslo, Norway\goodbreak
\and
Jet Propulsion Laboratory, California Institute of Technology, 4800 Oak Grove Drive, Pasadena, California, U.S.A.\goodbreak
\and
Max-Planck-Institut f\"{u}r Astrophysik, Karl-Schwarzschild-Str. 1, 85741 Garching, Germany\goodbreak
\and
Planetek Hellas, Leoforos Kifisias 44, Marousi 151 25, Greece\goodbreak
}

\authorrunning{Planck Collaboration}
\titlerunning{Diffuse component separation}

\abstract{We constrain polarized foreground emission between 30 and
  70\,GHz with the \textit{Planck} Low Frequency Instrument (LFI) and
  \textit{WMAP} data within the global Bayesian \BP\ framework. We
  combine for the first time full-resolution \textit{Planck} LFI
  time-ordered data with low-resolution \textit{WMAP} sky maps at 33,
  40 and 61\,GHz.  Spectral parameters are fit with a likelihood
  defined at the native resolution of each frequency channel. This
  analysis represents the first implementation of true
  multi-resolution component separation applied to CMB observations
  for both amplitude and spectral energy distribution (SED)
  parameters. For synchrotron emission, we approximate the SED as a
  power-law in frequency and find that the low signal-to-noise ratio
  of the current data strongly limits the number of free parameters
  that may be robustly constrained. We partition the sky into four
  large disjoint regions (High Latitude; Galactic Spur; Galactic
  Plane; and Galactic Center), each associated with its own power-law
  index. We find that the High Latitude region is prior-dominated,
  while the Galactic Center region is contaminated by residual
  instrumental systematics. The two remaining regions appear to be
  signal-dominated, and for these we
  derive spectral indices of $\beta_{\mathrm
    s}^{\mathrm{Spur}}=-3.17\pm0.06$ and $\beta_{\mathrm
    s}^{\mathrm{Plane}}=-3.03\pm0.07$, in good agreement with previous
  results. For thermal dust emission we assume a modified blackbody
  model and we fit a single power-law index across the full sky. We
  find $\beta_{\mathrm{d}}=1.64\pm0.03$, which is slightly steeper
  than reported from \textit{Planck} HFI data, but
  still statistically consistent at the 2\,$\sigma$ confidence level.}


\maketitle
 
\section{Introduction}
\label{sec:introduction}

One of the most important sources of information about the early universe is the
cosmic microwave background (CMB; \citealp{penzias:1965}). By mapping and
characterizing the statistical properties of this signal, cosmologists have
during the last few decades constrained both the composition and evolution of
the universe to percent accuracy \citep[e.g.,][]{bennett2012,planck2016-l01}.

As shown by the \COBE-FIRAS experiment \citep{mather:1994}, the frequency
spectrum of the CMB may to a very high precision be described in terms of a
blackbody with a mean temperature of $T_{\mathrm{CMB}}=2.7255\,\mathrm{K}$
\citep{fixsen2009}. As such, its peak intensity occurs at $161\,\mathrm{GHz}$,
and the primary frequency range considered by most CMB experiments is therefore
around 30 to 300\,GHz. In addition to the CMB, a diverse range of astrophysical
emission mechanisms contribute to the observed signal at these frequencies, both
of Galactic and extragalactic origin. For intensity, the main contributors
are synchrotron, free-free, CO, thermal dust, and anomalous microwave emission,
while for polarization, synchrotron and thermal dust emission dominate
\citep[e.g.,][and references therein]{bennett2012,planck2016-l04}.

At the foreground minimum, around 80\,GHz, the CMB anisotropies dominate over
the combined foreground amplitude over most of the sky \citep{planck2014-a10},
and CMB temperature extraction is therefore relatively straightforward. The same
does not hold true for polarization, even at the foreground minimum, the sum of
synchrotron and thermal dust emission is greater than the polarized CMB by
almost a full order of magnitude on large angular scales. Polarized foreground
estimation therefore plays a critically important role in contemporary
cosmology, as such observations may contain signatures of primordial
gravitational waves created during the Big Bang \citep[e.g.,][and references
therein]{kamionkowski:2016}, and thereby provide a unique observational window
on inflation and physics at the Planck energy scale.
  
However, the amplitude of the primordial gravitational wave signal is expected
to be smaller than 10--100\,nK on large angular scales \citep{tristram:2021}.
This, combined with a plethora of confounding instrumental effects, such as
temperature-to-polarization leakage \citep{bp12}, correlated noise \citep{bp06}
and calibration uncertainties \citep{bp07}, makes high-precision CMB
polarization science a particularly difficult challenge. Furthermore, as
summarized by the \Planck\ team in a document called ``Lessons learned from
\Planck'',\footnote{\href{URL}{https://www.cosmos.esa.int/web/planck/lessons-learned}}
the quality of current state-of-the-art CMB observations is limited by the
interplay between instrumental and foreground effects. This insight formed the
basis for the \BP\ project \citep{bp01}, which aims to implement an end-to-end
Bayesian CMB analysis framework that jointly accounts for both systematic
effects and astrophysical foregrounds, starting from raw time-ordered data. This
framework employs an explicit parametric model that accounts jointly for
cosmological, astrophysical, and instrumental parameters. These parameters are
sampled with Markov Chain Monte Carlo methods, such as Gibbs sampling, as implemented in
the \commander\ software. \citep{eriksen:2004,eriksen2008,bp03}.

The \BP\ results are described in a suite of 17 companion papers \citep[see][and
references therein]{bp01}, each focusing on a particular aspect of the analysis.
The current paper focuses on polarized foreground characterization, both in
terms of algorithms and results, with special attention paid to the spectral
properties of polarized synchrotron emission on large angular scales. The
current \BP\ analysis considers only the \Planck\ LFI observations in terms of
time-ordered data, although selected preprocessed external data sets are also
included to break critical degeneracies, specifically \WMAP\ measurements
between 33 and 61\,GHz \citep{bennett2012}, \Planck\ HFI measurements at 353 and
857\,GHz \citep{planck2016-l01,planck2020-LVII}, and the Haslam 408\,MHz
measurements \citep{haslam1982}, the latter two are only included in
temperature. Overall, the main emphasis of the present analysis lies on
frequencies below the foreground minimum, between 30 and 70\,GHz, and in
particular on the spectral behavior of polarized synchrotron and thermal dust
emission within this critically important frequency range. The present analysis
is the first to combine high-resolution \Planck\ measurements with
low-resolution \WMAP\ observations into a single coherent model, when estimating
both foreground amplitudes and spectral parameters.

The rest of this paper is structured as follows: In Sect.~\ref{sec:model}, we
briefly review the \BP\ data model, focusing on the aspects that are relevant
for polarized foreground analysis. In Sect.~\ref{sec:data}, we review the data
sets that are included in the current analysis. In Sect.~\ref{sec:methods} we
describe the basic algorithms, and connect these to the larger Gibbs sampling
framework outlined in \citet{bp01}. Results are presented in
Sect.~\ref{sec:results}, before we conclude in Sect.~\ref{sec:conclusions}.

\section{The \BP\ data model}
\label{sec:model}

As described by \citet{bp01}, the main goal of this project is to
perform end-to-end Bayesian CMB analysis, building on well-established
statistical methods. The first step in any such Bayesian analysis is
to write down an explicit parametric data model that will be
fit to observations through posterior mapping techniques. For \BP,
we adopt the following model for this purpose,
\begin{equation}
  \begin{split}
    d_{j,t} = g_{j,t}&\P_{tp,j}\left[\B^{\mathrm{symm}}_{pp',j}\sum_{c}
      \M_{cj}(\beta_{p'}, \Dbp^{j})a^c_{p'}  + \B^{\mathrm{asymm}}_{j}\left(s^{\mathrm{orb}}_{j}  
      + s^{\mathrm{fsl}}_{j}\right)\right] + \\
    + &s^{\mathrm{1hz}}_{j,t} + n^{\mathrm{corr}}_{j,t} + n^{\mathrm{w}}_{j,t},
  \end{split}
  \label{eq:todmodel}
\end{equation}
where $j$ represents a radiometer (or detector) label, $t$ indicates a
single time sample, $p$ denotes a single pixel on the sky, and $c$
represents one single astrophysical signal component. Further, 
$d_{j,t}$ denotes the measured time-ordered data; $g_{j,t}$ denotes
the instrumental gain; $\P_{tp,j}$ is a pointing matrix; $\B_{pp',j}$
denotes beam convolution; $\M_{cj}(\beta_{p}, \Dbp)$ denotes a
foreground mixing matrix that depends on some set of spectral
parameters, $\beta$, and instrumental bandpass specification, $\Dbp$;
$a^c_{p}$ is the amplitude of astrophysical component $c$ in pixel
$p$, measured at the same reference frequency as the mixing matrix
$\M$, and expressed in brightness temperature units;
$s^{\mathrm{orb}}_{j,t}$ is the orbital CMB dipole signal;
$s^{\mathrm{fsl}}_{j,t}$ denotes the contribution from far side-lobes;
$s^{\mathrm{1hz}}_{j,t}$ denotes the contribution from electronic 1\,Hz spikes;
$n^{\mathrm{corr}}_{j,t}$ denotes correlated instrumental noise; and
$n^{\mathrm{w}}_{j,t}$ is uncorrelated instrumental noise with
(diagonal) time-domain covariance matrix $\N^{\mathrm{w}}_j$. For
further details regarding any of these objects, we refer the
interested reader to \citet{bp01} and references therein.

The current paper focuses primarily on diffuse astrophysical
foregrounds, which for practical purposes are stationary in time,
and we are therefore not interested in the time-domain aspects of
the model. For this reason, we rewrite Eq.~\eqref{eq:todmodel} into
the following compact form,
\begin{equation}
  m_{\nu,p} = \left[\B^{\mathrm{symm}}_{pp'}\sum_{c}
    \M_{c}(\beta_{p'}, \Dbp)a^c_{p'}\right] + n^{\mathrm{w}}_{p},
  \label{eq:mapmodel}
\end{equation}
where $m_{\nu,p}$ is a binned sky map derived by co-adding all
radiometer data within a single frequency channel, and we are for the
moment conditioning on all time-domain parameters,
\begin{equation}
\left(\sum_{j \in \nu} \P_j^t (\N^{\mathrm{w}}_{j})^{-1} \P_j\right) \m_{\nu} =
\sum_j \P_j^t (\N^{\mathrm{w}}_j)^{-1}\r_j.
\label{eq:binmap}
\end{equation}
Here, $\r_j$ represents the cleaned and calibrated time-ordered data for
detector $j$, as defined by Eq.~(76) in \citet{bp01}, and full
marginalization over time-domain parameters is done iteratively
through Gibbs sampling; see Sect.~\ref{sec:gibbs} for further details.

In this paper, we are particularly interested in the total sky signal,
which may be written as follows,
\begin{equation}
  \s = \M\a \equiv \sum_{c=1}^{N_{\mathrm{comp}}}\a_c\, \left[\,U \int f_c(\nu; \beta)\,
    \tau(\nu)\,\mathrm d\nu\right].
  \label{eq:mixmat}
\end{equation}
Here, each astrophysical signal component, $c$, is associated with an
overall amplitude parameter $\a$; a unit conversion factor $U$ for
going from either thermodynamic or intensity units to brightness
temperature; a spectral energy density, $f_c$, describing the
intensity of the component relative to some reference frequency,
$\nu_{\mathrm{0}}$; and some set of spectral parameters, $\beta$,
which characterize the frequency dependence of the various emission
mechanisms. Finally, $\tau$ represents an instrumental bandpass
response function that describes the detector sensitivity as a
function of frequency. Again, we refer the interested reader to
\citet{bp01} for further details.

\subsection{Polarized sky model}

We assume in this paper that the polarized microwave sky may be well approximated by three
physically distinct components; synchrotron emission, thermal dust emission, and
CMB. A spinning dust, or anomalous microwave emission (AME) component is also
commonly included in similar studies, however with an upper limit of its
polarization fraction at $<$$1\,\%$ \citep{QUIJOTE_II_2016, macellari2011, bp15}, we
neglect it in this study. 

Each component exhibits a distinctly different behavior both in terms
of spatial structure and frequency dependence, and must be described
by some unique set of spectral parameters, $\beta$. The choice of
spectral parameters is both important and nontrivial. On the one hand,
it is important that the adopted parametric model for each component
is able to provide an acceptable goodness-of-fit, typically as
measured by some \chisq\ statistic. On the other hand, it is also
important that the model is not too flexible, as degeneracies between
components may increase the effective noise level to arbitrary high
levels. Generally speaking, degeneracies also tend to exacerbate
instrumental systematic errors, by attempting to accommodate small
signal discrepancies within the unconstrained parameters in the signal
model.  Choosing an appropriate parametric model is thus a delicate
trade-off between allowing sufficient flexibility to model the real
sky, while not introducing too many parameters, which otherwise may
increase both systematic and statistical uncertainties to untenable
levels.

For \BP, we adopt the following parametric model for the polarized
microwave sky, 
\begin{align}
\s_{\mathrm{RJ}} = &\,\a_{\mathrm{CMB}}\,
\frac{x^2e^{x}}{\left(e^{x}-1\right)^2} \frac{\left(e^{x_0}-1\right)^2}{x_0^2e^{x_0}} \,\\
+ &\,\a_{\mathrm{s}}\,
\left(\frac{\nu}{\nu_{0,\mathrm{s}}} \right)^{\beta_{\mathrm{s}}} \, \label{eq:synch}\\
+ &\,\a_{\mathrm{d}}\,\left(\frac{\nu}{\nu_{0,\mathrm{d}}}\right)^{\beta_{\mathrm{d}}+1}
\frac{e^{h\nu_{\mathrm{0,\mathrm{d}}}/kT_{\mathrm{d}}}-1}{e^{h\nu/kT_{\mathrm{d}}}-1}\label{eq:dust},
\end{align}
where all vectors are expressed in brightness (Rayleigh-Jeans) temperature
units; $x=h\nu/kT_0$ (where $h$ and $k$ are Planck's and Boltzmann's constants
respectively and $T_0$ is the CMB monopole of 2.7255\,K; \citealp{fixsen2009}); and $\nu_{0,c}$ is the
reference frequency for component $c$, $\a$ is its amplitude, $\beta_{\mathrm s}$ is the
synchrotron power-law index, and $\beta_{\mathrm d}$ and $T_{\mathrm d}$ are the thermal dust
power-law index and temperature, respectively. Even this minimal model, which adopts a 
power-law SED for synchrotron and modified blackbody SED for thermal dust,
contains far too many parameters to be fit per pixel with the \BP\ data set, and several
informative priors will be imposed to regularize the system, as discussed in
detail below.

\subsubsection{Synchrotron emission}
\label{sec:synch_model}

\begin{figure}[t] 
\center 
\includegraphics[trim=0 250 0 250,width=\linewidth]{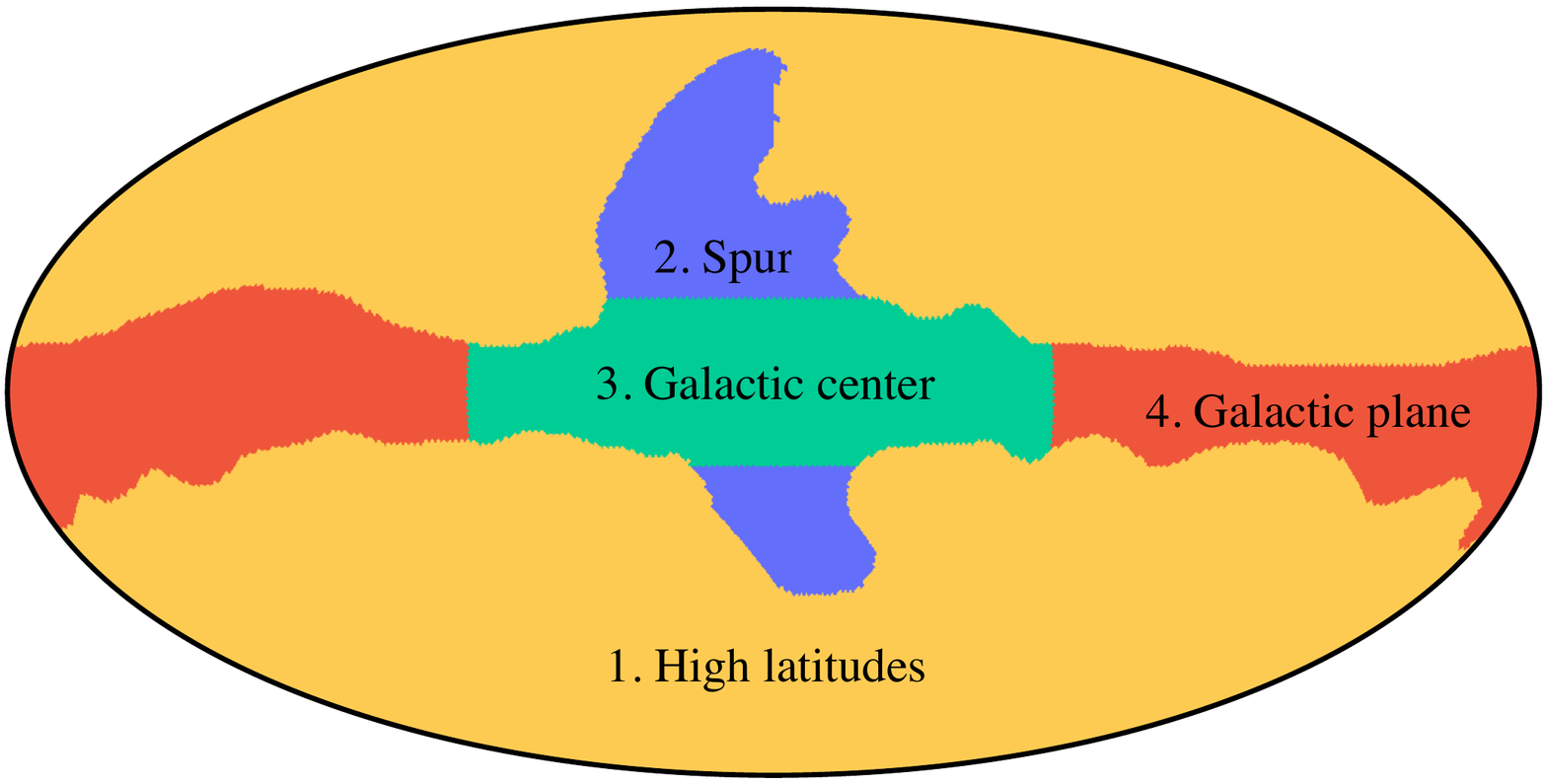}
\includegraphics[width=\linewidth]{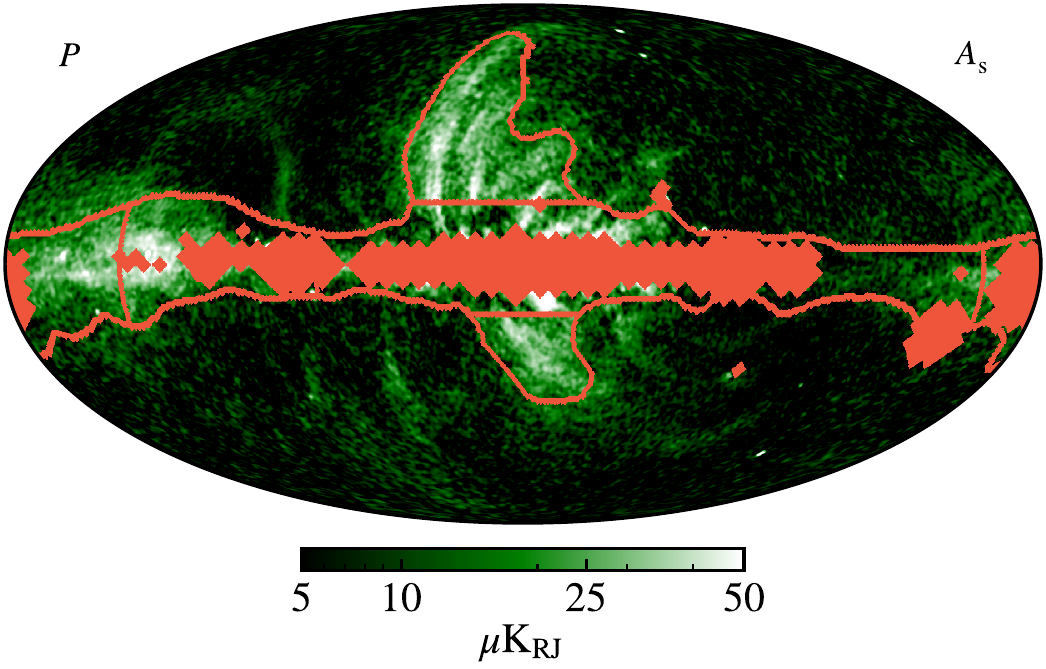}\\ 
\includegraphics[width=\linewidth]{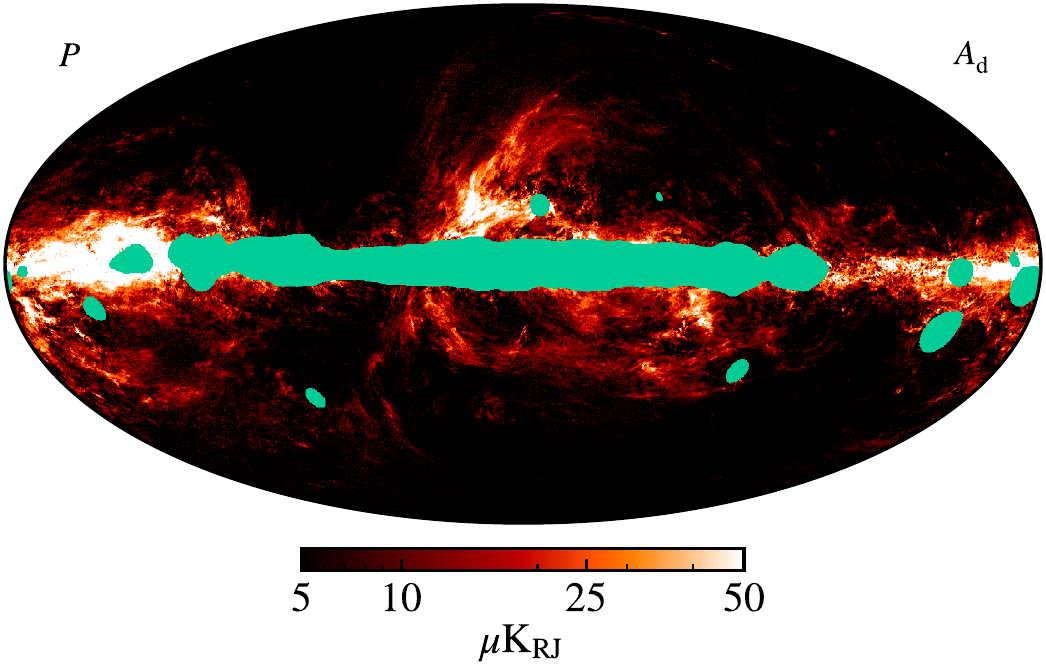}\\ 
\caption{\emph{Top:} Final region map used in the sampling procedure
  for the synchrotron spectral index. \emph{Middle:} \Planck\ 2018
  polarized synchrotron amplitude map with the corresponding spectral
  index processing mask for \BP\ and a region outline that divides the sky into
  six disjoint regions. This was further reduced to four, motivated
  primarily by the low signal-to-noise apparent in some of the
  regions. The mask is chosen to reduce temperature-to-polarization
  leakage bias, and to exclude bright point sources. \emph{Bottom:}
  \Planck\ 2018 polarized thermal dust amplitude with the
  corresponding processing mask.}\label{fig:regions}
\end{figure}

The first foreground component, and the brightest Galactic emission mechanism
between 30 and 70\,GHz, is synchrotron radiation. This emission is generated
when relativistic cosmic-ray electrons ejected from supernovae gyrate in the
Galactic magnetic field. From both temperature and polarization observations,
the synchrotron SED is known to be well approximated by a power-law over several
decades in frequency \citep[e.g.,][]{lawson1987, reich1988, Platania2003,
Davies2006, gold2009}, at least from 1\,GHz to 100\,GHz, although some analyses
also claim evidence for a slight spectral steepening towards higher frequencies
\citep{kogut:2012,jew2019}. The \texttt{Galprop} model \citep{orlando:2018} is a
physically motivated model that takes into account quantities such as the
electron temperature ($T_e$) and large-scale magnetic field distributions, and
this physical model also predicts SED flattening below 1\,GHz depending on the
model parameters. However, because the flattening occurs at frequencies well
below 10\,GHz, this is not relevant for the current analysis, which only
considers frequencies above 30\,GHz, and we therefore assume a straight
power-law SED model for synchrotron emission in the following.

Next, several analyses have reported significant spatial variations in
$\beta_\mathrm{s}$
\citep[e.g.,][]{fuskeland2014,krachmalnicoff2018,fuskeland:2019}. In
general, these analyses report flatter spectral indices (around
$\beta_{\mathrm{s}}=-2.8$) at low Galactic latitudes, and steeper at
high Galactic latitudes (around $\beta_{\mathrm{s}}=-3.1$). This
picture appears roughly consistent with similar results derived from
intensity observations \citep[e.g.,][]{vidal2014, platania1998,
lawson1987, reich1988}. On the other hand, when analyzing
low-resolution \WMAP\ polarization data with full correlated noise
propagation, \citet{dunkley2009} found only minor differences between
low and high Galactic latitudes.

\begin{table*}
  \newdimen\tblskip \tblskip=5pt
  \caption{Overview of all included data bands in the \BP\ polarization analysis. The columns represent the following; 1) Science experiment for that band, 2) frequency band name, 3) data processing pipeline 4) $\healpix$ resolution in $\nside$, 5) highest included multipole $\ell_{\mathrm{max}}$, 6) beam resolution in arcminutes, 7) bandpass center frequency in GHz, 8) bandpass width in GHz, 9) noise rms given in $\muK$ scaled by pixel size, 10) noise format.}
  \label{tab:data}
  \vskip -4mm
  \footnotesize
  \setbox\tablebox=\vbox{
   \newdimen\digitwidth
   \setbox0=\hbox{\rm 0}
   \digitwidth=\wd0
   \catcode`*=\active
   \def*{\kern\digitwidth}
    \newdimen\dpwidth
    \setbox0=\hbox{.}
    \dpwidth=\wd0
    \catcode`!=\active
    \def!{\kern\dpwidth}
    \halign{\hbox to 2.cm{#\leaderfil}\tabskip 1.5em&
      \hfil$#$\hfil \tabskip 1.5em&
      \hfil$#$\hfil \tabskip 1.5em&
      \hfil$#$\hfil \tabskip 1.5em&
      \hfil$#$\hfil \tabskip 1.5em&
      \hfil$#$\hfil \tabskip 1.5em&
      \hfil$#$\hfil \tabskip 1.5em&
      \hfil$#$\hfil \tabskip 1.5em&
      \hfil$#$\hfil \tabskip 1.5em&
      #\hfil\tabskip 0em\cr
  \noalign{\doubleline}
  \omit&\omit&\omit&\omit&\omit&\omit\hfil\sc Resolution\hfil&\nu_{c}&\omit\hfil\sc Bandwidth\hfil& \sigma &\omit\cr
  \omit\hfil\sc Experiment\hfil&\omit\hfil\sc Band\hfil&\omit\hfil\sc Processing\hfil& \nside & \ell_{\mathrm{max}} & [\mathrm{arcmin}]& \omit\hfil[GHz]\hfil&\omit\hfil[GHz]\hfil&[\muK\,\mathrm{arcmin}]&\omit\hfil\sc Noise Format\hfil\cr
  \noalign{\vskip 3pt\hrule\vskip 5pt}
  \Planck\ LFI & *30 & \BP\    & *512 & 1500 & 32\farcm4*&*28.4&*5.7&260& TOD+White noise\cr
  \Planck\ LFI & *44 & \BP\    & *512 & 2000 & 27\farcm1*&*44.1&*8.2&370& TOD+White noise\cr
  \Planck\ LFI & *70 & \BP\    & 1024& 2500 & 13\farcm3*&*70.1&14.0&270& TOD+White noise\cr
  \Planck\ HFI & 353& \Planck\ \mathrm{DR4} & 1024& 3000 & *4\farcm94&353!*&88.2&380& White noise\cr
  \WMAP       & \textit{Ka} & \WMAP & **16&**64& 40\arcm**&*33!*&*7.0&410& Full covariance\cr 
  \WMAP       & \textit Q  & \WMAP & **16&**64& 31\arcm**&*41!*&*8.3&380& Full covariance\cr
  \WMAP       & \textit V  & \WMAP & **16&**64& 21\arcm**&*61!*&14.0&460& Full covariance\cr
  \noalign{\vskip 5pt\hrule\vskip 5pt}}}
  \endPlancktablewide
  \par
  \end{table*}

As shown in the following, the \BP\ data combination\footnote{Note that the \BP\
data combination does not include the \WMAP\ \textit K-band sky, as this channel would
otherwise compete with \Planck\ LFI 30\,GHz in terms of total signal-to-noise
ratio.} has in general low statistical power to probe spatial variations. We
therefore partition the sky into four large disjoint regions, as shown in the
top panel of Fig.~\ref{fig:regions}, and fit one spectral index for each
region. For reference, the middle panel in this figure shows the \Planck\ 2018
synchrotron amplitude map \citep{planck2016-l04} with region outlines and a
corresponding processing mask used in \BP. 

Since the spectral index is fit uniformly over large sky regions, it is
important to impose a processing mask during the posterior evaluation to avoid
pixels with poor goodness-of-fit from contaminating surrounding pixels. To this
end, we construct a dedicated spectral index processing mask for synchrotron
emission by first fitting the component amplitudes given our chosen prior. We
then smooth these map to $4\deg$ FWHM, and construct a corresponding mask by
thresholding on the strongest 5\,\% of the signal to remove the brightest part
of the Galactic plane. Next, we combine the resulting mask with a smoothed
\chisq\ map thresholded at the strongest 10\,\% of the signal to remove strong
sources such as Tau A and their surrounding area.  Finally, we use an
un-smoothed \chisq\ map in order to remove particularly bright compact sources.
The resulting mask is visualized along with the region outlines in the middle
panel of Fig.~\ref{fig:regions}.

The region set was defined as follows: We started from the 24-region
partitioning defined by \citet{fuskeland2014}, which itself was based
on the nine year \WMAP\ polarization analysis mask
\citep{bennett2012}. We then ran preliminary analyses to determine the
effective posterior width resulting from each region. If this was
found to be strongly prior-dominated, we merged neighboring
regions. This process resulted in four main regions, which we will
refer to as 1) High Latitudes; 2) the Galactic Spur or simply Spur; 3)
the Galactic Center; and 4) the Galactic plane. As we will show later,
even after this process the high-latitude region does not have
sufficient signal-to-noise ratio to significantly constrain
$\beta_{\mathrm s}$. The reason for this is illustrated in the middle
panel of Fig.~\ref{fig:regions}; the synchrotron emission in this
region is generally very faint, and there is very little leverage to
estimate spectral variations as a function of frequency.

We also note that synchrotron emission is sensitive to Faraday
rotation \citep[e.g.,][]{beck:2013} in areas with high electron
densities and magnetic fields, which is the case near the Galactic
center. This effect is caused by circular birefringence, where left-
and right-handed circularly polarized emission traverse the magnetic
field of an ionized medium at different velocities, generating a net
linear polarization signal with a polarization angle proportional to
the field strength. However, since this effect is proportional to the
squared wavelength of the radiation, it is most prominent at
frequencies below 5\,GHz, and it is negligible for most of the sky
above 30\,GHz. In this paper, we do not make any corrections for
Faraday rotation to any frequency channel. However, as shown in
Sect.~\ref{sec:results}, our synchrotron constraints for the Galactic
center region are contaminated by systematic effects, and
these are most likely due to a combination of instrumental and
astrophysical mis-modeling effects.

\subsubsection{Thermal dust emission}

The second most significant polarized foreground component between 30
and 70\,GHz is thermal dust emission generated by interstellar dust
grains that collectively account for $\approx$$\,1\,\%$ of the mass of
the interstellar medium; see, e.g., \citet{hensley2020} for a recent
review of relevant physics and observations constraints. The size of
these grains typically varies from a few angstroms to a few tenths of
a $\mu\mathrm{m}$, and they are heated up by the interstellar
radiation field to a temperature of about 20\,K. This heat is then
re-emitted thermally with a peak frequency between 1000 and 2000\,GHz
(as defined in brightness temperature units). Furthermore, due to
paramagnetic dissipation resulting from interactions between rotating
grains and the local magnetic field, the dust grains tend to align
with their short axes parallel to the local magnetic field. In turn,
this alignment can induce a polarization signal with a polarization
fraction of 20\,\% or more, as reported by \citet{planck2016-l11A}.

The most commonly used thermal dust SED model in CMB studies is that of a modified blackbody
function, as defined in Eq.~\eqref{eq:dust}. This SED has three free
parameters; 1) the amplitude $\a_{\mathrm d}$, which traces the
surface density of dust particles along each line-of-sight; 2) a
spectral index, $\beta_{\mathrm d}$, that quantifies the low-frequency
slope of the SED, and depends on the physical composition of the dust
grains; and 3) the dust temperature, $T_{\mathrm d}$, which affects
the peak location of the modified blackbody function, and depends on
the local radiation field strength at any given position. More
complex, and perhaps more realistic, representations do exist,
including multi-component modified blackbody
\citep[e.g.,][]{finkbeiner1999, meisner2015} or physical dust grain
models \citep[][]{guillet2018}. 

The particular data combination used in the \BP\ analysis has a very low
signal-to-noise ratio for constraining thermal dust SED variations, whether of
frequency or spatial origin, as we are working at synchrotron dominated frequencies.
In this paper, we therefore only consider one single free parameter for the thermal
dust SED, namely a spatially constant value of $\beta_{\mathrm{d}}$. The bottom
panel in Fig.~\ref{fig:regions} shows the \Planck\ thermal dust amplitude map
\citep{planck2016-l04} with its corresponding processing mask used in the
sampling procedure. This mask was generated by thresholding the $10\,\%$ largest
values of the smoothed \chisq\ map as described in the last section.  The thermal
dust temperature map is fixed pixel-by-pixel to that derived from \Planck\
temperature observations between 30 and 857\,GHz by \citet{planck2020-LVII}.

No attempts are made to account for SED variations along each line of
sight. As emphasized by \citet{Tassis2015}, the sum of two modified
blackbodies is not a modified blackbody, and spatial variations in
either $\beta_{\mathrm{d}}$ or $T_{\mathrm{d}}$ along each
line-of-sight will therefore necessarily break the current model. This
effect is particularly important in polarization, since the magnetic
field direction also varies along the line-of-sight, potentially
resulting in different SEDs for the two Stokes parameters, $Q$ and
$U$. However, these variations are far too weak to be measured with
the current data set.

\section{Data selection}
\label{sec:data}

As described by \citet{bp01}, a primary motivation for the \BP\ project is to
establish a common analysis platform for past, current, and future CMB
observations that supports end-to-end analysis, from raw time-ordered data to
final high-level products such as astrophysical component maps and cosmological
parameters. To support and guide the algorithm development process, the \Planck\
LFI observations were chosen as a first real-world application for this
framework, a choice that was primarily motivated by the fact that this data set
is already well known to the collaboration members, and secondarily because of its modest
data volume and relatively benign instrumental systematic effects.

At the same time, it is clear that the LFI data are by themselves not able to
constrain all relevant astrophysical components. There are at least five
significant diffuse components between 30 and 70\,GHz; CMB, synchrotron,
free-free, AME, and thermal dust emission, while the LFI data only comprise
three independent frequency channels. The LFI data must therefore be augmented
with external observations in order to derive a statistically non-degenerate
model. In principle, the \Planck\ HFI data \citep{planck2016-l03} would be an
ideal match, providing strong constraints on CMB, free-free and thermal dust
emission. However, if these data were to be included the \BP\ analysis in their
entirety, they would dominate over the LFI observations in terms of total
sensitivity, which  would undermine the main purpose of the current
presentation, which focuses on the algorithms themselves. For this reason, we
choose to include only the \Planck\ HFI 857\,GHz channel in temperature and the
353\,GHz channel in polarization, to constrain the amplitude of thermal dust
emission. In both cases, we adopt the latest rendition of these maps published
by the \Planck\ team, as derived by the \Planck\ Data Release 4 (DR4) pipeline,
also known as \textit{NPIPE} \citep{planck2020-LVII}.

The same argument applies to the \WMAP\ \textit K-band channel
\citep{bennett2012}. While this channel provides strong constraints on both AME
and polarized synchrotron emission, its statistical sensitivity is so high that
it would rival that of the \Planck\ LFI 30\,GHz channel if included in the \BP\
analysis, which again would undermine the main purpose of the current work. For
this reason, we include only the \WMAP\ \textit{Ka}-, \textit Q-, and \textit
V-band channels in the following, centered on 33, 41 and 61\,GHz, respectively.\footnote{The \textit W-band channel is omitted because it is known to be contaminated by
systematic residuals; see \citet{bennett2012,bp17} for details.}

In addition, we include the Haslam 408\,MHz survey to constrain synchrotron
emission in temperature, see \citet{bp13}. This results in a total of eight frequency channels in
intensity, and seven frequency channels in polarization, which in
principle should be sufficient to constrain a model with five intensity
components and three polarization components.

One of the important novel algorithmic developments that will be described in
Sect.~\ref{sec:methods} is true multi-resolution component separation for both
linear and non-linear foreground parameters. Specifically, we consider in the
following only the low-resolution \WMAP\ polarization data at a
HEALPix\footnote{\url{http://healpix.jpl.nasa.gov}} \citep{gorski2005}
resolution of $N_{\mathrm{side}}=16$, for which dense pixel-pixel noise
covariance matrices are available \citep{bennett2012}. This ensures that we have
at least nominally a complete noise description for all frequency channels
relevant for CMB extraction, and also that small-scale polarization results are
entirely dominated by \Planck\ LFI.

All data except the \Planck\ DR4 353\,GHz channel are analyzed as provided
by the respective team without any preprocessing or smoothing. The
353\,GHz channel, however, is smoothed to an angular resolution of
$10\arcm$ and re-pixelized at a resolution of
$N_{\mathrm{side}}=1024$, corresponding to a pixel size of
$3\farcm4$. This is done both in order to speed up the analysis
process \citep{bp01,bp03}, and to suppress small-scale correlated
noise which would otherwise lead to a significant
\chisq\ excess. Table~\ref{tab:data} provides an overview over all
data sets included in the current analysis.

\section{Methods}
\label{sec:methods}

We now aim to fit the data model in Eq.~\eqref{eq:todmodel} to
the data listed in Table~\ref{tab:data}. We define
$\omega=\{g,n_{\mathrm{corr}},\Dbp,\a,\beta,\ldots\}$ to be the set of
all free parameters in this model of both instrumental and
astrophysical origin. Our main goal is then to map out the
corresponding joint posterior distribution, which is given by Bayes'
theorem,
\begin{equation}
    P(\omega\mid \d) = \frac{P(\d\mid \omega)P(\omega)}{P(\d)} \propto
  \mathcal{L}(\omega)P(\omega).
  \label{eq:jointpost}
\end{equation}
In this expression, $P(\omega\mid \d)$ is the posterior distribution, and
describes our knowledge about $\omega$ after performing the experiment in
question; $\mathcal{L}(\omega)\equiv P(\d\mid \omega)$ is called the likelihood,
and quantifies the information content in $\d$ regarding $\omega$; and
$P(\omega)$ is called the prior, which quantifies our beliefs regarding $\omega$
before doing the experiment. The prior can also be used actively to regularize
specific degeneracies in the model.

This posterior distribution is large and complex with billions of free
parameters. Attempting to directly map out the full posterior
distribution by brute-force is infeasible. Instead, we resort to
Markov Chain Monte Carlo methods, and draw samples from the posterior
distribution. In particular, we find that the Gibbs sampling algorithm
\citep{geman:1984} is particularly well suited to handle the
complexity of this model. We therefore adopt the \commander\ CMB Gibbs
sampling framework as the starting point for our analysis. This
software was first introduced by \citet{eriksen:2004} for optimal CMB
power spectrum estimation applications, building on ideas originally suggested by
\citet{jewell2004} and \citet{wandelt2004}, and later generalized to
also account for joint CMB and component separation by
\citet{eriksen2008} and \citet{seljebotn:2019}. In this operational
mode, \commander\ played a central role in the official
\Planck\ analysis, as summarized by
\citet{planck2013-p01,planck2014-a01,planck2016-l01} and references
therein. \BP\ has now generalized this framework further to also
account for low-level TOD processing and mapmaking, effectively
turning the entire CMB analysis challenge into one global problem.

\subsection{Gibbs sampling}
\label{sec:gibbs}

Gibbs sampling formalizes the idea of iterative analysis within a rigorous
statistical language. In short, the theory of Gibbs sampling states that samples
from a complex joint distribution may be drawn by iteratively sampling from each
(typically simpler) conditional distribution. The full \BP\ Gibbs chain
\citep{bp01} illustrates how this is done in practice, and can be written as
\begin{alignat}{11}
\g &\,\leftarrow P(\g&\,\mid&\,\d,&\, & &\,\xi_n, &\,\Dbp, &\,\a, &\,\beta, &\,C_{\ell})\\
\n_{\mathrm{corr}} &\,\leftarrow P(\n_{\mathrm{corr}}&\,\mid&\,\d, &\,\g, &\,&\,\xi_n, &\,\Dbp, &\,\a, &\,\beta, &\,C_{\ell})\\
\xi_n &\,\leftarrow P(\xi_n&\,\mid&\,\d, &\,\g, &\,\n_{\mathrm{corr}}, &\,&\,\Dbp, &\,\a, &\,\beta, &\,C_{\ell})\\
\Dbp &\,\leftarrow P(\Dbp&\,\mid&\,\d, &\,\g, &\,\n_{\mathrm{corr}}, &\,\xi_n, &\,&\,\a, &\,\beta, &\,C_{\ell})\\
\a &\,\leftarrow P(\a&\,\mid&\,\d, &\,\g, &\,\n_{\mathrm{corr}}, &\,\xi_n, &\,\Dbp, &\,&\,\beta, &\,C_{\ell})\label{eq:amps}\\
\beta &\,\leftarrow P(\beta&\,\mid&\,\d, &\,\g, &\,\n_{\mathrm{corr}}, &\,\xi_n, &\,\Dbp, &\,\a,&\,&\,C_{\ell}) \label{eq:specind}\\
C_\ell &\,\leftarrow P(C_\ell&\,\mid&\,\d, &\,\g, &\,\n_{\mathrm{corr}}, &\,\xi_n, &\,\Dbp,&\,\a, &\,\beta, &\,)&.
\end{alignat}
Here, the symbol $\leftarrow$ means setting the variable on the left-hand side
equal to a sample from the distribution on the right-hand side. For convenience,
we also define the notation ``$\omega\setminus \xi$'' to imply the set of
parameters in $\omega$ except $\xi$.

The parameters not already introduced are noise power spectrum
parameters ($\xi_n$),
bandpass corrections ($\Dbp$), and the CMB power spectrum
($C_{\ell}$). Since the current paper focuses on polarized
foregrounds, we are here primarily interested in the amplitude
parameters, $\a$, sampled in Eq.~\eqref{eq:amps}, and the spectral
parameters, $\beta$, sampled in Eq.~\eqref{eq:specind}. We will
therefore describe these two sampling steps in detail in the
following, and we refer the interested reader to \citet{bp01} and
references therein for details regarding the remaining steps.

\subsection{Signal amplitude sampling}
We first consider the signal amplitude conditional distribution,
$P(\a\mid\d, \omega\setminus\a)$, which has already been a primary
focus of interest for a long line of studies, including
\citet{jewell2004}, \citet{eriksen:2004, eriksen2008}, and
\citet{seljebotn:2019}. In the following, we give a brief summary of
these developments, and also highlight two important novel features
introduced in the current work.

The appropriate starting point for this conditional distribution is the data
model described in Eq.~\eqref{eq:mapmodel}. We first note that $\a$ contains all
signal amplitude maps, using some appropriate linear basis, stacked column-wise,
such that  $\a = [\a_{\mathrm{CMB}}, \a_{\mathrm{s}}, \a_{\mathrm{d}}]^t$, where
the $t$ superscript denotes the transpose operator. Further, we adopt a spherical
harmonics basis to describe each diffuse component, such that $\a_i =
\{a^T_{i,\ell m},a_{i,\ell m}^E,a_{i,\ell m}^B\}$ contains the various
temperature and polarization spherical harmonics coefficients
\citep{zaldarriaga1997} for component $i$. Second, for notational convenience we
combine the beam and mixing matrix operators in Eq.~\eqref{eq:mapmodel} into one
joint linear operator, $\A_{\nu}\equiv\B_{\nu}\M_{\nu}$, such that this equation
may be written succinctly as
\begin{equation}
  \m_{\nu} = \A_{\nu}\a + \n_{\nu}.
  \label{eq:simple_data_model}
\end{equation}
Noting that the noise, $\n_{\nu} = \m_{\nu} - \A_{\nu}\a$, is
assumed to be Gaussian distributed with vanishing mean and a known covariance 
matrix, $\N_{\nu}$, Bayes' theorem then allows us to write the conditional
distribution of interest in the following form,
\begin{align}
  P(\a\mid\d, \omega\setminus\a) &\propto P(\d\mid\omega)P(\a)\\
  &\propto P(\m\mid\a)P(\a)\\
  & \propto
  \left(\prod_{\nu}\mathrm \exp\left(-\frac{1}{2}(\m_{\nu}-\A_{\nu}\a)^t\N_{\nu}^{-1}(\m_{\nu}-\A_{\nu}\a)\right)\right)\nonumber\\
  &\cdot \mathrm \exp \left(-\frac{1}{2}(\a-\bar{\a})^t\S_{\nu}^{-1}(\a-\bar{\a})\right).
  \label{eq:posterior}
\end{align}
Here, the second line holds because TOD binning is a deterministic operation when
conditioning on $\omega$ \citep{bp01}, and the set of binned sky maps, $\m$, is
a sufficient statistic for $\a$; no additional information in the TOD can possibly
provide more knowledge about the foreground amplitude maps beyond that stored in
$\m$ if $\omega$ is exactly known. In the last line we adopt a Gaussian signal prior with mean $\bar{\a}$ and
covariance matrix $\S$, discussed further below. In this paper, we set
$\bar{\a}=0$, and only use $\S$ for smoothing purposes. Since the product of two
Gaussian distributions is another Gaussian, $P(\a\mid\d, \omega\setminus\a)$ is
also Gaussian, and the appropriate sampling algorithm is given by a standard
multivariate Gaussian, which is discussed in detail in Appendix~A in
\citet{bp01}. In particular, a proper sample may be drawn by solving the
following linear equation for $\a$,
\begin{equation}
  \biggl(\S^{-1} + \sum_{\nu}\A^t_{\nu}\N_{\nu}^{-1}\A_{\nu}\biggr)\,\a = 
  \sum_\nu\A_{\nu}^t\N_\nu^{-1}\m_{\nu} + \sum_{\nu}\A_{\nu}^t\N_{\nu}^{-1/2}\eta_{\nu} +
  \S^{-1/2}\eta_{0},
\label{eq:ampl_samp_wiener}
\end{equation}
where $\eta_\nu$ and $\eta_0$ are random Gaussian vectors of independent $N(0,
1)$ variates. Because of its large dimensionality, this equation must in
practice be solved using iterative linear algebra techniques, and we use a
preconditioned conjugate gradient solver for this purpose \citep[e.g.,][]{shewchuk:1994}.

This exact equation was discussed in detail by \citet{seljebotn:2019},
who also introduced two novel and efficient preconditioners for
partial-sky observations based on pseudo-inverse and multi-grid
techniques. However, in the present work, for which no analysis mask
is imposed during the main Gibbs sampling analysis, we adopt the
simpler diagonal pre-conditioner described by \citet{eriksen2008},
which both converges slightly faster with full-sky data, and has a
lower computational cost per iteration. The two main novel features
regarding Eq.~\eqref{eq:ampl_samp_wiener} introduced by \BP\ (as
described by \citet{bp13} for temperature and by this paper for
polarization) concern the data and model representation and active use
of spatial priors.

\subsubsection{Vector representation and object-oriented programming}

Equation ~\eqref{eq:ampl_samp_wiener} is written in a general vector
form, without reference to any specific vector representation or
basis, or to any specific form of either the effective mixing matrix,
$\A_{\nu}$, or noise covariance matrix, $\N_{\nu}$. All of these may,
in principle, be chosen per component and frequency channel. As
detailed by \citet{bp03}, we have implemented such flexibility in the
most recent version of \commander\ adopting an object-oriented
programming style, where separate classes are defined for each
object. For instance, the current implementation contains specific
classes for three general types of components, namely diffuse
components (for which $\a$ is represented in terms of spherical
harmonic coefficients, $a_{\ell m}$), point source objects (for which
$\a$ is represented in terms of a single flux density per compact
source), and fixed spatial templates (for which $\a$ is represented in
a single multiplicative amplitude per template). Any combination of
such objects may all be fit simultaneously and jointly through
Eq.~\eqref{eq:ampl_samp_wiener}. It is also relatively straightforward
to add new types of objects as the need may arise. Two examples of
classes that might be important for future applications include pixel-
or needlet-based components \citep[e.g.,][]{marinucci:2008}, which
could be useful for modeling partial sky experiments.

Each of the instrumental objects are also implemented in terms of individual
classes. This is particularly relevant for the noise covariance matrix,
$\N_{\nu}$, which may have very different representations for different
experiments. For example, in the current analysis, we model the \Planck\ LFI
noise as a sum of correlated and white noise, where the correlated noise is
treated as a stochastic variable in the Gibbs chain, and sampled over directly,
while the white noise uncertainty is propagated analytically through a diagonal
$\N_{\nu}$. This approach supports, for the first time, propagation of both
correlated and white noise at all angular resolutions. For \WMAP, however, we do
not yet have access to time-ordered data within our framework, so $\N_{\nu}$ is
defined in terms of the precomputed dense low-resolution noise covariance
matrices provided by the \WMAP\ team \citep{bennett2012}, which is
computationally feasable because this data is smoothed to $\nside=16$. For this
reason, \WMAP\ contributes only to large angular scales in the current analysis.
Finally, for the \Planck\ 353\,GHz measurements, which are essential for
modeling polarized thermal dust emission at full angular resolution, we are for
now only able to propagate white noise uncertainties with a diagonal $\N_{\nu}$
as given by \citet{planck2020-LVII}. Of course, this will necessarily lead to an
underestimation of dust-related uncertainties; hence modeling \Planck\ HFI
observations in time-domain is clearly a high-priority issue for future work,
but outside the scope of the current project.

In summary, the first critically important novel feature provided by
the current \commander\ implementation is the ability to operate with
fundamentally different types of data sets within one analysis and
thereby exploit complementary features from each data set to break
degeneracies within the full model. At the time of publication, there
is direct support for only three types of noise covariance matrices,
namely diagonal matrices, \WMAP-style dense Stokes $QU$ matrices, and
diagonal matrices with explicit marginalization over low-$\ell$
modes. However, using the new infrastructure described here and by
\citet{bp03}, it is straightforward to add support for other types of
matrices. Possibly useful examples include block-diagonal or banded
covariance matrices, or noise covariance matrices with specific
subspaces projected out through operator based filters. The latter
could be particularly useful for ground-based experiments that tend to
have limited statistical sensitivity on large angular scales, but high
systematic uncertainties. We hope that such features can be
implemented, and made publicly available, by third-party authors as
Open Source contributions \citep{bp05}.

\subsubsection{Gaussian spatial priors}
\label{sec:priors}

The second novel feature supported by the latest \commander\ implementation is
the use of active spatial priors for the diffuse components. For practical
purposes, we currently support only Gaussian priors, as described by
Eq.~\eqref{eq:posterior}, as non-Gaussian priors would lead to a prohibitively
high computational cost associated with the current conjugate gradient-based approach. In
practice, imposing a spatial prior is therefore equivalent to specifying a prior
mean, $\bar{\a}$, and covariance matrix, $\S$, for each component. In the current
polarization-oriented analysis, however, we do not wish to enforce any
informative priors on any of the three free components
(polarized CMB, synchrotron or thermal dust emission), so we set
$\bar{\a}=0$ for all three. Instead, we use $\S$ to adjust the allowed
level of fluctuations around zero for each component as a function of angular
scale; this is numerically equivalent to choosing an appropriate effective
smoothing scale for each component, and might for this paper be considered more of a technical
issue than a prior in the normal sense. For an example of an application of
active priors, however, see \citet{bp13}, in which proper informative spatial
priors are imposed on both free-free and anomalous microwave emission.

The remaining question is how to choose a specific form of $\S$ for each
component. There are two main requirements for this choice. First, $\S$ is the
covariance of $\a$, and therefore dictates the overall fluctuation level of the
fitted components. A large value of $\S$ implies a weak prior, and the fitted
component will then be dominated by the data-driven likelihood term in
Eq.~\eqref{eq:posterior}, while a low value of $\S$ implies a strong prior,
resulting in values close to the prior mean. In practice, we want the prior to
play a limited role where the data have a large signal-to-noise ratio, but a
stronger role where the data are noise dominated. It is therefore in general
desirable to choose scale-dependent priors, in which the smoothing becomes
gradually stronger; the prime example of this is a standard Gaussian smoothing
kernel.

As already mentioned, we adopt spherical harmonics as our basis set
for $\a$, writing each diffuse polarized signal component as
\begin{equation}
a_{X}(\hat{n}) = \sum_{\ell m} a^{X}_{\ell m} Y_{\ell m}(\hat{n}).
\end{equation}
For each component we therefore also define an angular power spectrum prior of the form
\begin{equation}
  \hat{D}^{X}_{\ell} = \left<|a^{X}_{\ell m}|^2\right> \,\ell(\ell+1)/2\pi,
\end{equation} 
where $X=\{E,B\}$. Note that this power spectrum
prior representation is conceptually similar to the assumptions made by \smica\
\citep{cardoso2008}, one of the four CMB extraction codes used by \Planck.

We adopt the following prior variances for each of the three
astrophysical components in the current analysis,
\begin{align}
  \hat{D}^{-1}_{\mathrm{CMB}}(\ell) &= 0\,\mu\mathrm{K}^2\\
  \hat{D}_{\mathrm{s}}(\ell) &= 200\, \mathrm e^{-\ell(\ell+1)\sigma^2(30\arcm)}
  \,\mu\mathrm{K}^2\\
  \hat{D}_{\mathrm{d}}(\ell) &= 500\, \mathrm e^{-\ell(\ell+1)\sigma^2(10\arcm)}
  \,\mu\mathrm{K}^2,
\end{align}
where $\sigma^2(\theta_{\mathrm{FWHM}})\equiv (8\ln 2)\,
\pi/(180\cdot60)\theta_{\mathrm{FWHM}}^2$ is the standard deviation of a
Gaussian distribution expressed in terms of a full width at half maximum,
$\theta_{\mathrm{FWHM}}$, in arcminutes.

Note that for the CMB component, the inverse variance is set to zero,
which is equivalent to an infinite variance---or simply no prior at
all. For the polarized synchrotron and thermal dust amplitudes, the
priors correspond to Gaussian smoothing of $30\arcm$ and $10\arcm$
FWHM, respectively. The overall scaling factors are chosen to be 200
and $500\,\muK^2$, respectively, which correspond to the observed
angular power spectrum of each component on large angular scales. In
these cases, the priors are used to apodize the resulting foreground
amplitude maps with Gaussian smoothing kernels, to avoid ringing
around bright sources and the Galactic plane and unphysical
degeneracies at high multipoles.

Future analyses may wish to use $\S$ to directly estimate the
angular power spectrum of each foreground component, fully analogous to the CMB
case \citep{bp11,bp12}. This will then both alleviate the need of specifying the
prior parameters by hand before executing the analysis, and it will ensure
optimal smoothing properties for each component, resulting in minimal
high-$\ell$ degeneracies between the various components.

Finally, we conclude this section by emphasizing that the above priors are in
fact only priors, and not deterministic smoothing operators. Thus, the resulting
synchrotron and thermal dust component maps will be determined by the properties
of the observed data wherever the data are stronger than the prior.
This is important to bear in
mind for instance when trying to estimate the angular power spectrum of the
resulting maps; the component maps $\a$ that result from solving
Eq.~\eqref{eq:ampl_samp_wiener} correspond to a model of the sky without
instrumental beam convolution, but with spatially varying noise properties,
depending on the local signal-to-noise ratio of the data. The angular
resolutions of the synchrotron and thermal dust maps are not given
precisely by a Gaussian beam of 30 and 10\arcm, respectively, but will generally
be higher where the data are sufficiently strong. As such, the behavior of these
foreground maps is conceptually similar to the \texttt{GNILC} algorithm
\citep{Remazeilles2011b}, in which a spatially varying angular resolution also
results from signal-to-noise ratio variations. Likewise, it is also important to
note that the noise properties of these maps are highly non-trivial, and the
only statistically robust way of assessing and propagating their uncertainties
is through the ensemble of sky map samples produced by the algorithm itself.

\subsection{Spectral parameter sampling}

The second of the two main conditional distributions discussed in this paper is
$P(\beta\mid\d, \omega\setminus\beta)$, which describes the foreground SED
parameters. In general, $\a$ and $\beta$ are strongly correlated for a given
component, especially for high signal-to-noise data. For temperature-oriented
foreground analysis, the \BP\ Gibbs sampler therefore implements a
special-purpose sampling step for these parameters, by exploiting the definition
of a conditional distribution, $P(\a,\beta\mid\d) =
P(\a\mid\d,\beta)P(\beta\mid\d)$ \citep{stivoli:2010,bp01,bp13}. That is, we
first sample spectral parameters from the marginal distribution with respect to
$\a$, and then sample $\a$ conditionally with respect to $\beta$. The resulting
algorithm is thus effectively an independence sampler in $\{\a,\beta\}$ (i.e., a
sampler where the new proposal does not depend on the previous) with an internally
vanishing Markov chain correlation length. In this case, any long-term Markov
chain correlations come from degeneracies with other parameters.

However, as described by \citet{stivoli:2010}, this algorithm is only
computationally practical for observations with the same angular resolution, as
the computational expense for evaluating the marginal distribution $P(\beta\mid\d)$
otherwise becomes prohibitively high. In practice, this means that all data maps
must be smoothed to a common angular resolution. For temperature data, this is
not a major problem, but for polarization analysis it is non-trivial, since the
smoothing operation correlates the instrumental noise. In addition, the data
combination used in the current paper involves low-resolution \WMAP\ data with
dense noise covariance matrices, and smoothing all data to this resolution is
impractical.

Since we focus on polarization in this work, we instead adopt a
standard Metropolis-within-Gibbs sampler for $\beta$. In this case,
the appropriate conditional posterior distribution may be derived from
Eq.~\eqref{eq:simple_data_model} by noting that the instrumental noise
is assumed to be Gaussian distributed with covariance matrix
$\N_{\nu}$, and recalling that the amplitude is for the moment assumed to
be perfectly known. Therefore,
\begin{equation}
  \begin{aligned}[t]
  P(\beta\mid\d, \a) &\propto P(\d\mid\a, \beta) P(\beta)\label{eq:beta_posterior}\\
  \propto \Bigg[\prod_{\nu} &
    \mathrm \exp\left(-\frac{1}{2}\left(\d_{\nu}-\A_{\nu}(\beta)\a\right)^t\N_{\nu}^{-1}\left(\d_{\nu}-\A_{\nu}(\beta)\a\right)\right)\Bigg]P(\beta),
  \end{aligned}
\end{equation}
where $\d=\{\d_{\nu}\}$ is the set of all available frequency maps (which within
the larger \BP\ framework may be a specific set of frequency map sky samples),
and $P(\beta)$ is a user-defined prior, typically a Gaussian distribution with
physically motivated mean and standard deviation.

Note that Eq.~\eqref{eq:beta_posterior} is also written with a general vector
notation, and makes no reference to a specific vector basis for either $\a$ or
$\beta$; it therefore allows the various data sets to be defined in
\emph{different} basis sets. The only requirement is that there must exist a
well-defined mapping between $\beta$ and the effective mixing matrix at each
frequency, $\A_{\nu}(\beta)$. This generality is precisely what is needed to
analyze multi-resolution observations jointly, for instance high-resolution LFI
data together with low-resolution \WMAP\ data.

As noted in Sect.~\ref{sec:synch_model}, the signal-to-noise ratios of the
\Planck\ and \WMAP\ polarization data are modest, so we have defined
broad regions for $\beta_{\mathrm{s}}$, as shown in Fig.~\ref{fig:regions}. To
sample these values, we employ a standard Metropolis algorithm, as outlined in
Algorithm~\ref{al:mh}, using Eq.~\eqref{eq:beta_posterior} as target
distribution and a standard symmetric Gaussian proposal distribution,
${T(\beta^{(i)}\mid\beta^{(i-1)})\sim N(\beta^{(i-1)},\C_\beta)}$, where $\C_{\beta}$
is a tunable proposal matrix. When evaluating Eq.~\eqref{eq:beta_posterior}, we
first propose $\beta_{\mathrm{s},i}$ independently for each region $i$, and then
smooth the resulting map with a $10^{\circ}$ FWHM Gaussian beam to suppress edge
effects. This smoothed map is then used to
evaluate the mixing matrix $\A_{\nu}$ at the appropriate resolution for each
frequency channel.

To assess the goodness-of-fit locally across the sky, we compute a \chisq\ map
per pixel. Since \WMAP\ is pixelized with ${N_{\mathrm{side}}=16}$, the \chisq\
map is too. The sum of this map, after application of an optional analysis mask,
is used to evaluate the exponent in Eq.~\eqref{eq:beta_posterior} and the
corresponding Metropolis acceptance rate.

\begin{algorithm}[t]
Initialize $\beta^{(0)} \sim P(\beta)$\\
\For{i=$1$, n}{
  Propose new index map: $\beta^{\mathrm{prop}} \sim N\left(\beta^{(i-1)},\C_{\beta}\right)$ \\
  Compute acceptance rate:\quad $\alpha=\min \left\{1, \frac{P(\beta^{\mathrm{prop}}\mid\d,\a)}{P(\beta^{(i-1)}\mid\d,\a)}\right\}$\\
  Draw $u \sim$ Uniform$(0,1)$\\
  \eIf{$u<\alpha$}{
    Accept the proposal: $\beta^{(i)} \leftarrow \beta^{\mathrm{prop}}$
    }{
    Reject the proposal: $\beta^{(i)} \leftarrow \beta^{(i-1)}$
    }
}
\caption{Metropolis algorithm used for spectral index sampling.}
\label{al:mh}
\end{algorithm}

As summarized in Algorithm~\ref{al:mh}, this method has two free tunable
parameters, the proposal matrix $\C_\beta$, and the number of Metropolis
steps per main Gibbs iteration, $n$. Before starting a full Gibbs analysis, we
perform a tuning run using a diagonal $\C_\beta$ and adjust the
diagonal elements until the resulting accept rate is between 0.2 and 0.8. Once
that happens, we run a longer chain with fixed $\C_{\beta}$ and replace this
matrix with the covariance matrix from the resulting samples. Finally, we run a third
chain, computing the Markov autocorrelation length from the resulting chain,
and setting $n$ such that the empirical correlation between samples $\beta^{(i)}$
and $\beta^{(i+n)}$ is less than 0.1. These tuning steps are only performed once
for each analysis setup, and files are stored on disk for subsequent runs.

\subsection{Spectral index priors}

The only missing part of the algorithm is now a specification of the
priors. For the synchrotron and thermal dust spectral indices, we
adopt for the main analysis Gaussian priors of
$\beta_{\mathrm{s}}=-3.3\pm0.1$ and $\beta_{\mathrm{d}}=1.56\pm0.10$,
respectively. The former of these are motivated by the \Planck\ LFI
2018 likelihood analysis, which finds a linear scaling factor of
$\alpha=0.058\pm0.004$ between 30 and 70\,GHz. Accounting for the
bandpasses of the respective channels \citep{planck2016-l02,bp09}, the
quoted scaling factor translates into a spectral index of
$\beta_{\mathrm{s}}=-3.3$. The thermal dust spectral index prior is
also informed by the official \Planck\ analysis
\citet{planck2014-a12}, and in this case the \BP\ data sets do not add
any useful independent information to complement the original work,
because of the absence of the high signal-to-noise ratio HFI
measurements.

Many alternative synchrotron priors were explored during the course of
the project, varying both the mean and width. In general, these all
led to more significant residuals than the final choice. Regarding the
prior width, we note that $\sigma_{\beta}=0.1$ corresponds to only
$1\,\sigma$, and the full $3\,\sigma$ confidence interval therefore
spans from $-3.6$ to $-3.0$, all of which will be covered during the
full Gibbs run. As discussed by \citet{bp15}, shallower indices than
$\beta=-3.0$ are very difficult to accommodate with both LFI and
\WMAP\ without introducing an additional low-frequency dust correlated
component, for instance polarized AME. Consequently, a broader prior
width of $\sigma_{\beta}=0.2$, which permits synchrotron indices as
shallow as $\beta_{\mathrm{s}}=-2.7$, leads to large foreground
excesses that only can be accommodated by the CMB component within
the current model. The result is an obvious foreground bias in the
large-scale CMB extraction and a corresponding excess in estimates of
the optical depth of reionization \citep{bp12}. For further
discussions regarding different priors, see Sect.~\ref{sec:priors}.

We treat the four regions differently in terms of synchrotron
priors. While the Spur and Galactic plane regions are fitted using
both the likelihood and prior as described above, we only include the
prior for the high-Galactic and Galactic Center regions. This is
because of very strong degeneracies with respect to instrumental
systematic effects in these regions. Specifically, the Galactic Center
region is particularly susceptible to bandpass mismatch effects, as
the bandpass leakage corrections for the 30\,GHz channel are of order
unity in this region \citep{bp09}, while the High Latitude region is
particularly sensitive to relative gain uncertainties and degeneracies
with the CMB quadrupole \citep{bp07}. To prevent potential residual
systematic errors from contaminating these two regions, we only
marginalize over the prior in these cases. While clearly non-ideal, we
consider this approach to be preferable to simply fixing the
synchrotron index at some given value, as was done for most previous
\Planck\ polarization analyses, e.g.,
\citet{planck2014-a10}. Ultimately, this is a concession that the
current data set are not sufficiently strong to uniquely and
independently determine both synchrotron and CMB components without
priors, and additional measurements from high sensitivity
low-frequency experiments such as C-BASS \citep{jew2019} and QUIJOTE
\citep{QUIJOTE_I_2015} will be extremely valuable to further improve
the quality of the LFI and \WMAP\ data sets in the future.

\section{Results}
\label{sec:results}
\subsection{Goodness-of-fit}

\begin{figure}[t]
\center
\includegraphics[width=\linewidth]{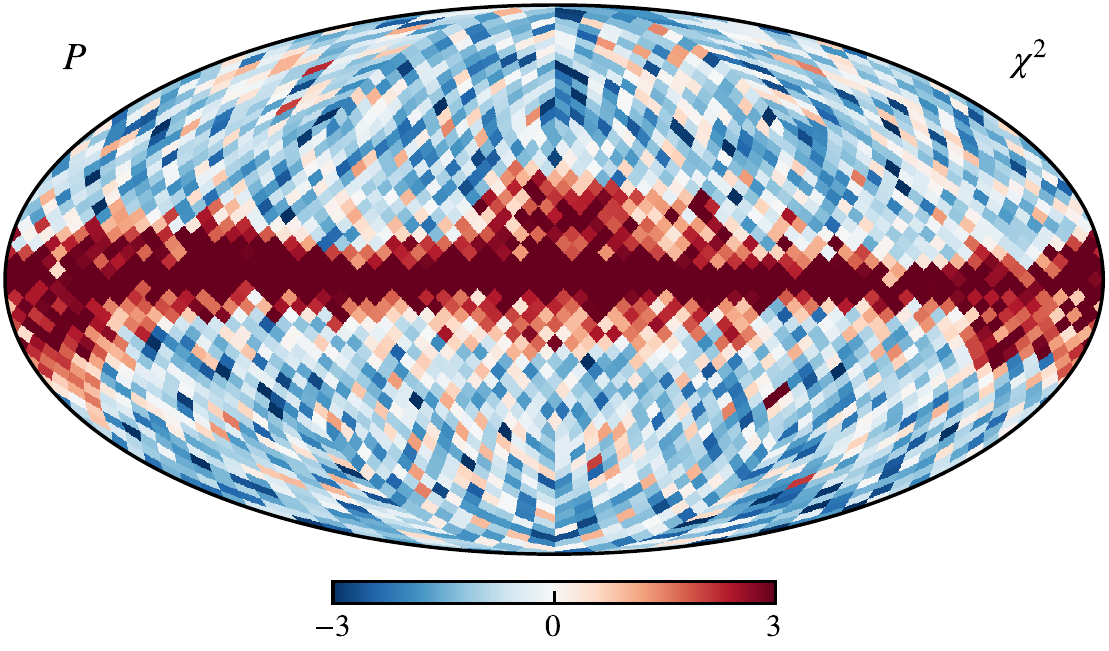}
\caption{Normalized and reduced \chisq\ per pixel, summed over Stokes
  $Q$ and $U$ and averaged over all Gibbs samples.}
\label{fig:chisq} 
\end{figure}

\begin{figure*}[t]
\center
\includegraphics[width=0.23\linewidth]{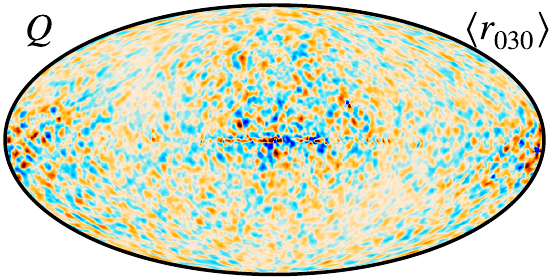}
\includegraphics[width=0.23\linewidth]{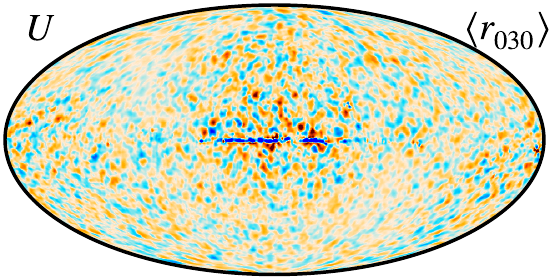}\hspace*{5mm}
\includegraphics[width=0.23\linewidth]{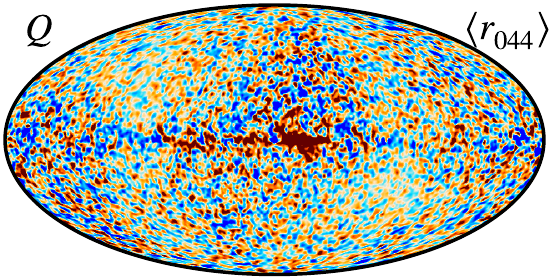}
\includegraphics[width=0.23\linewidth]{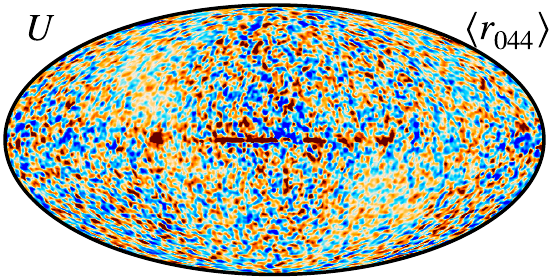}\\
\includegraphics[width=0.23\linewidth]{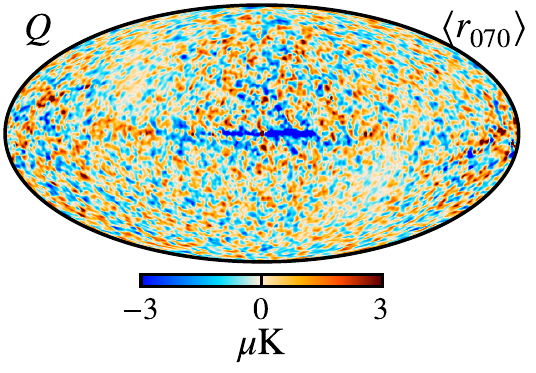}
\includegraphics[width=0.23\linewidth]{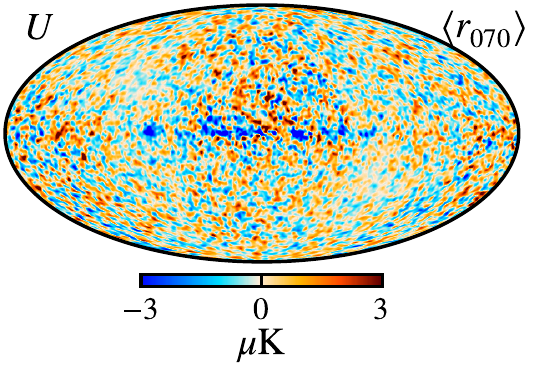}\hspace*{5mm}
\includegraphics[width=0.23\linewidth]{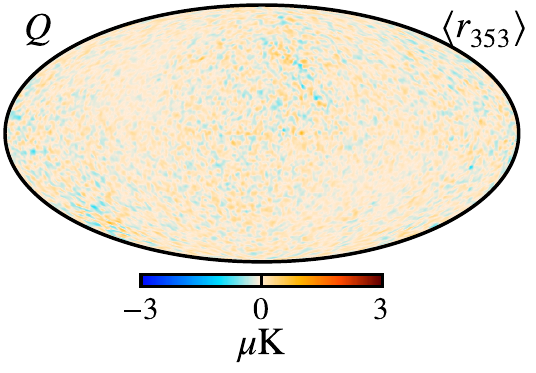}
\includegraphics[width=0.23\linewidth]{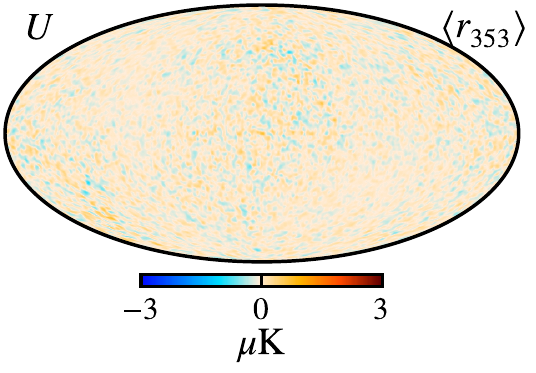}
\caption{Posterior mean total data-minus-model residual maps
($\r_{\nu}=\d_{\nu}-\s_{\nu}$) in $Q$ and $U$ for \BP\ LFI 30\,GHz (\textit{top
left}), 44\,GHz (\textit{top right}), 70\,GHz (\textit{bottom left}) and 353\,GHz
(\textit{bottom right}). All maps are smoothed to a common angular resolution of
$2^{\circ}$ FWHM.} 
\label{fig:resplanck}
\vspace*{5mm}
\includegraphics[width=.23\linewidth]{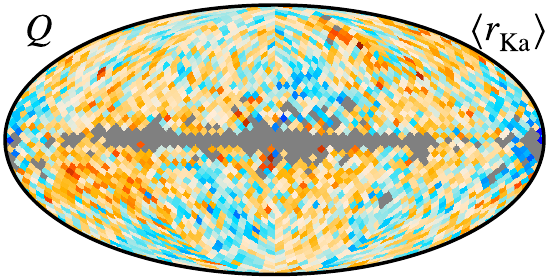}
\includegraphics[width=.23\linewidth]{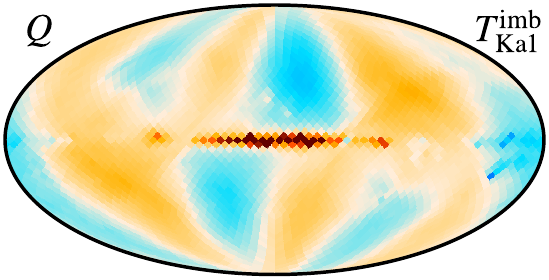}\hspace*{5mm}
\includegraphics[width=.23\linewidth]{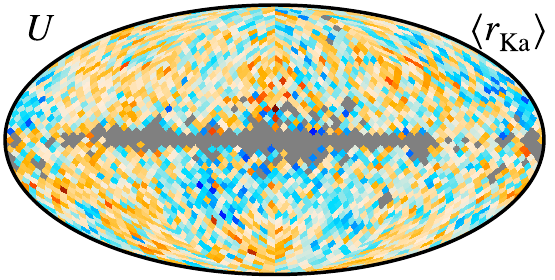}
\includegraphics[width=.23\linewidth]{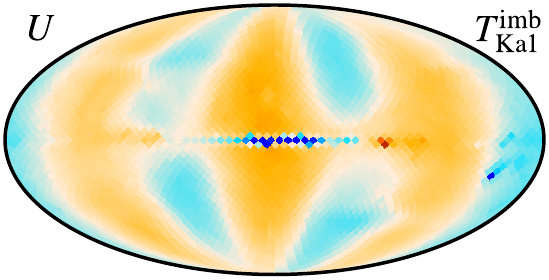}\\
\includegraphics[width=.23\linewidth]{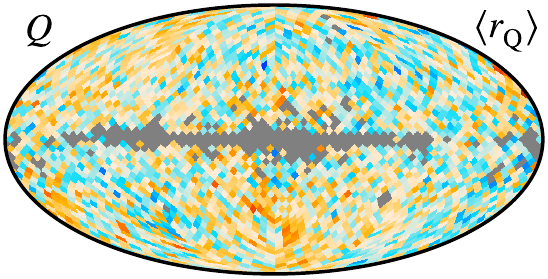}
\includegraphics[width=.23\linewidth]{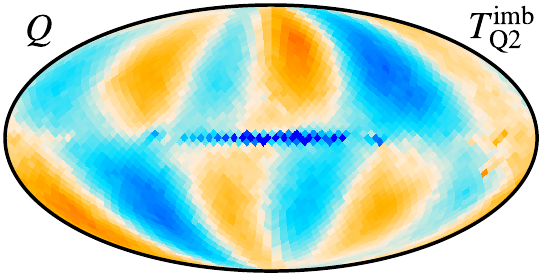}\hspace*{5mm}
\includegraphics[width=.23\linewidth]{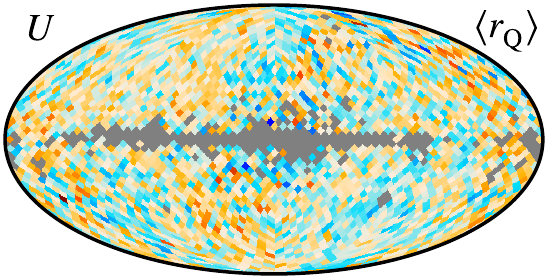}
\includegraphics[width=.23\linewidth]{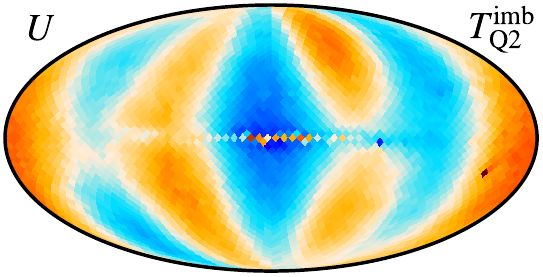}\\
\includegraphics[width=.23\linewidth]{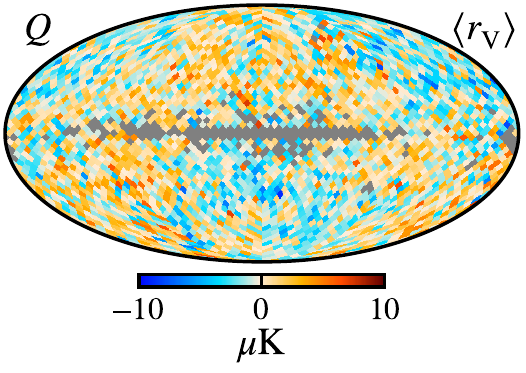}
\includegraphics[width=.23\linewidth]{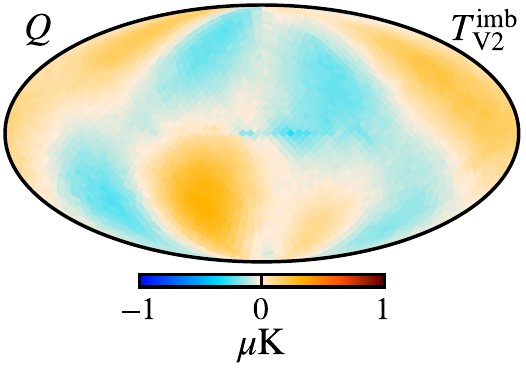}\hspace*{5mm}
\includegraphics[width=.23\linewidth]{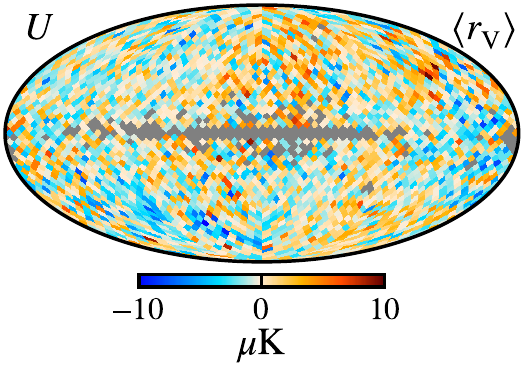}
\includegraphics[width=.23\linewidth]{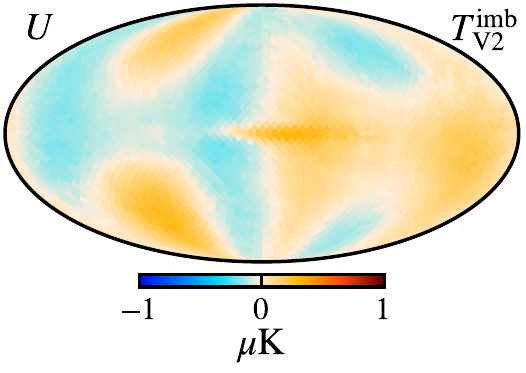}
\caption{Posterior mean total data-minus-model residual maps
($\r_{\nu}=\d_{\nu}-\s_{\nu}$) in $Q$ (\emph{first column}) and $U$ (\emph{third
column}) for all included \WMAP\ bands in the two left-most columns, at
$\nside=16$, masked with the processing mask applied in the pipeline. The second
and fourth columns show corresponding transmission imbalance template maps for
one of the differencing assemblies (eg. Q\{1,2\}, V\{1,2\}) per residual as derived by
\citet{jarosik2007}; note that these templates are each associated with an
unknown scaling amplitude that may take either sign.}
\label{fig:reswmap}
\end{figure*}

Before turning our attention to the final astrophysical products, we
assess the goodness-of-fit of the fitted sky model. Our first
statistic is the total \chisq\ per pixel, summed over frequencies and
Stokes $Q$ and $U$ parameters and averaged over Gibbs samples, as
shown in Fig.~\ref{fig:chisq}. To aid visual interpretation, this map
is plotted in the form
$(\chi^2-n_{\mathrm{dof}})/\sqrt{2n_{\mathrm{dof}}}$, where
$n_{\mathrm{dof}}=20\,798$ is an estimate of the total number of
degrees of freedom (i.e., number of full-frequency data pixels summed
over frequencies, minus the number of fitted parameters.) within each
$N_{\mathrm{side}}=16$ pixel. If the model performs as expected, this
quantity should have zero mean and unit standard deviation. Overall,
we see that our model appears to perform well at high Galactic
latitudes,\footnote{We note that it is very difficult to compute
  $n_{\mathrm{dof}}$ rigorously due to the presence of active priors,
  since each prior-constrained model parameter only contributes with a fraction of a
  degree-of-freedom. As such, the important feature of
  Fig.~\ref{fig:chisq} is its spatial structure, not the absolute
  zero-level. The uniform and slightly negative bias at high Galactic
  latitudes is thus simply an indication that $n_{\mathrm{dof}}$ is
  very slightly over-estimated; if it were due to over-estimating the
  white noise rms level, which is the only other possible explanation
  for a $\chi^2$ deficit, the \Planck\ scanning strategy would have
  been visually apparent.} while the Galactic plane shows clear excess
\chisq. This behavior is not surprising, as the Galactic Center is far
more complex and difficult to model, both in terms of astrophysics and
instrumental effects. We do note that the morphological structure of
the excess \chisq\ appears to be correlated with Galactic emission,
rather than instrumental effects.

\begin{figure*}[t]
\center 
\includegraphics[width=.49\linewidth]{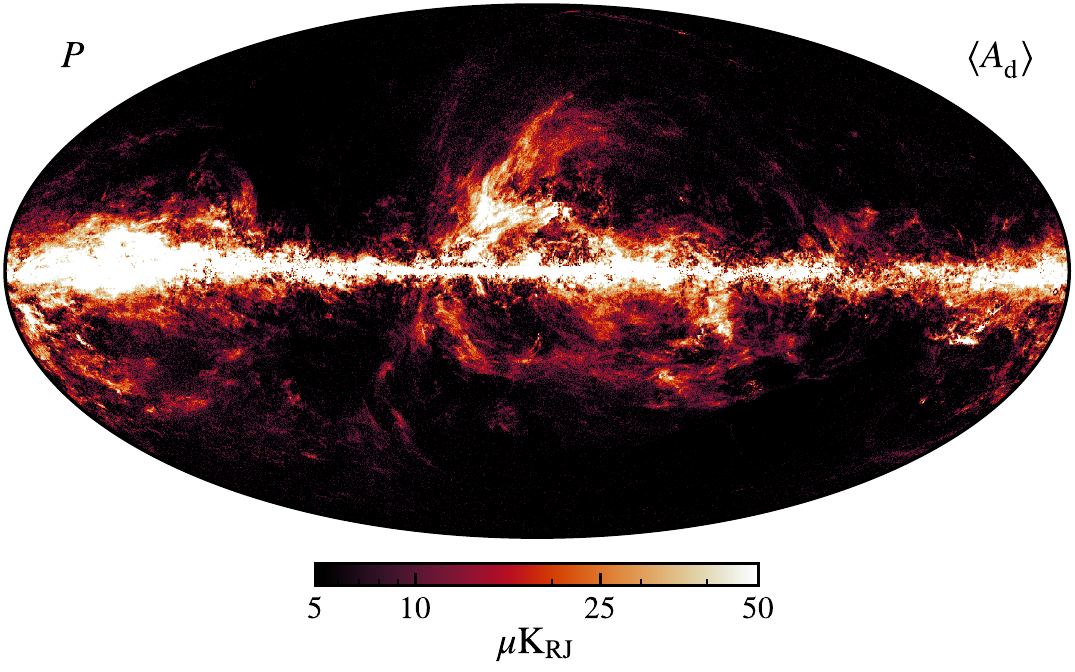}
\includegraphics[width=.49\linewidth]{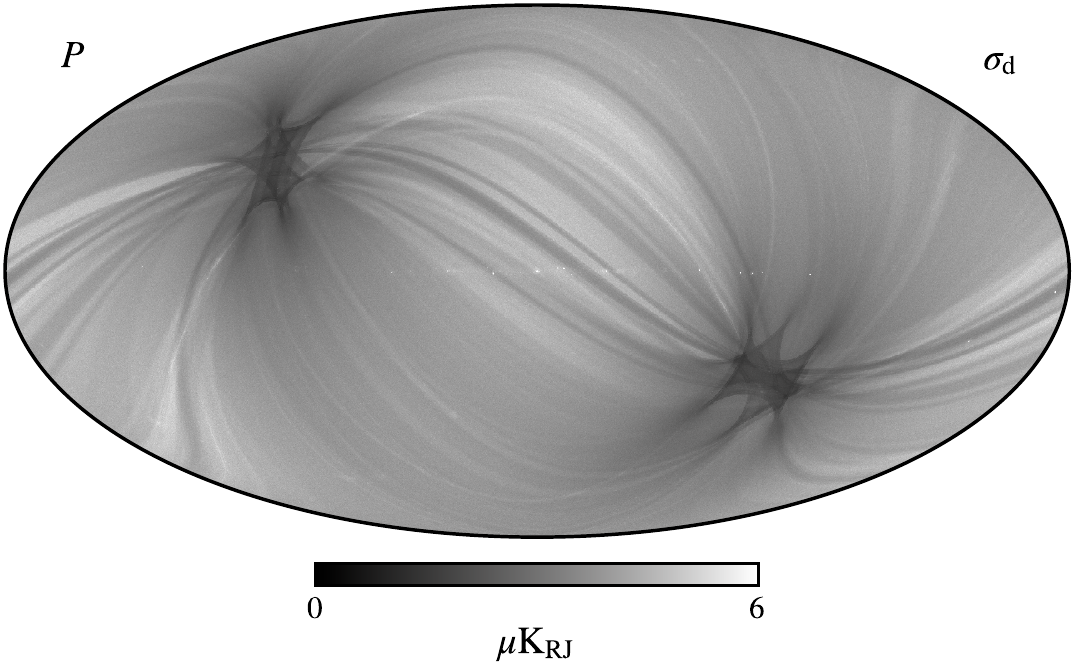}\\
\includegraphics[width=.49\linewidth]{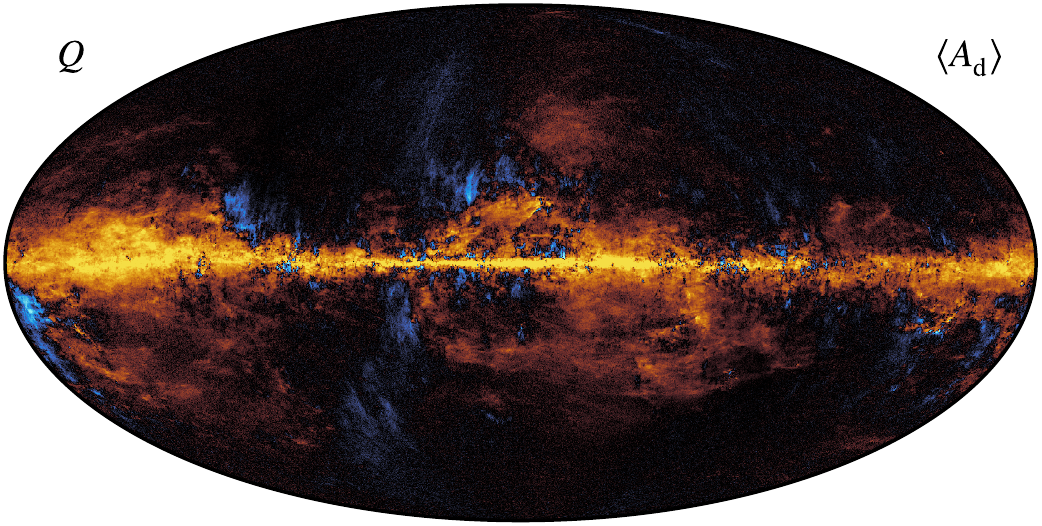}
\includegraphics[width=.49\linewidth]{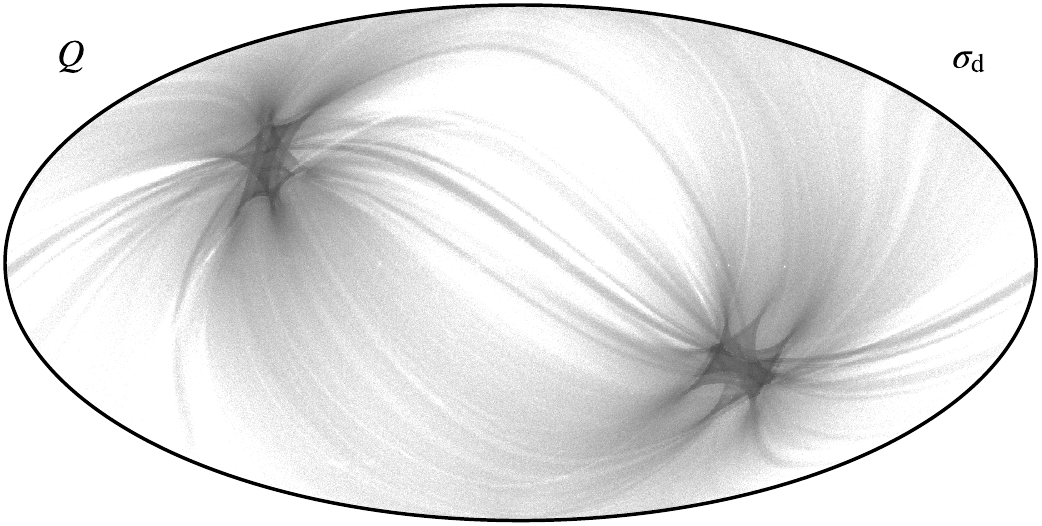}\\
\includegraphics[width=.49\linewidth]{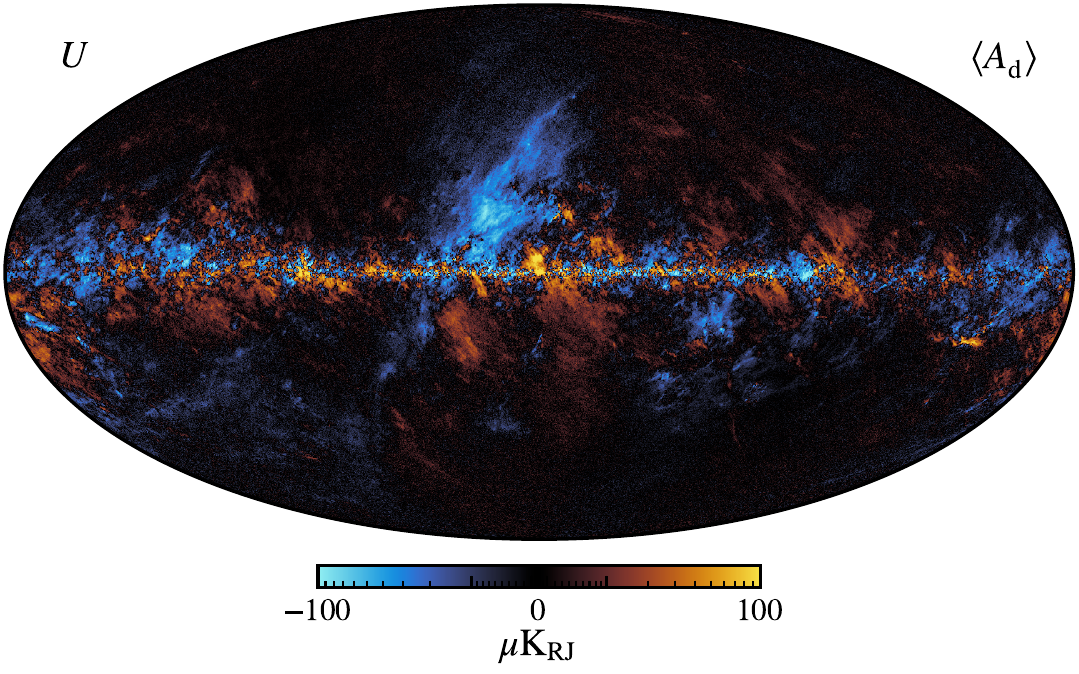}
\includegraphics[width=.49\linewidth]{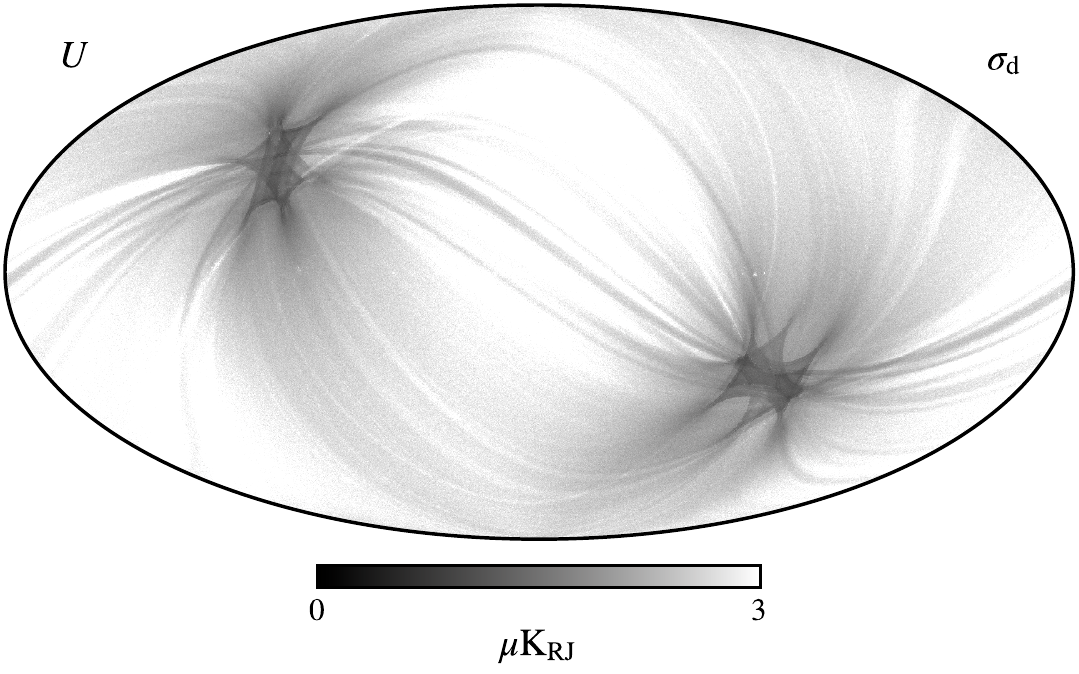}
\caption{(\emph{Left column}:) Polarized thermal dust amplitude maps,
  plotted in terms of the polarization amplitude $P=\sqrt{Q^2+U^2}$
  (\textit{top row}) and the Stokes $Q$ (\textit{middle row})
  and $U$ parameters (\textit{bottom row}).
  (\emph{Right column}:) Corresponding posterior standard
  deviation maps. All maps are evaluated at a reference frequency of
  353\,GHz in units of $\muK_{\mathrm{RJ}}$, and averaged over all
  available Gibbs samples. The effective angular
  resolution is $10\arcmin$ FWHM.}\label{fig:Ad}
\end{figure*}

Another useful statistic for evaluating the model goodness-of-fit is
data-minus-model residual maps per frequency,
${\r_{\nu}=\d_{\nu}-\s_{\nu}}$. These residual maps provide a visual
summary of remaining systematics on a band-to-band basis, which is
useful when physically interpreting specific artefacts in the
\chisq\ map. Figure~\ref{fig:resplanck} shows such residual maps for
the \Planck\ frequency bands, smoothed to $2\deg$ FWHM to suppress
white noise. Generally, these maps indicate excellent model
performance, with amplitudes of $\lesssim\,$$3\muK$. High
Galactic latitudes appear consistent with noise, while small
deviations are seen around the Galactic Center, with a morphology that
may suggest residual bandpass-induced temperature-to-polarization
leakage \citep{bp12}.

The 44\,GHz channel exhibits the largest relative variations. This is expected
from algorithmic arguments, by noting that this channel does not dominate the
determination of any single fitted astrophysical component. In contrast, the
30\,GHz channel dominates synchrotron determination; the 70\,GHz channel
dominates CMB determination; and the 353\,GHz channel dominates thermal dust
determination. As such, any excess fluctuations in these channels will rather be
interpreted as signal belonging to whichever foreground component that uses this
as a reference band. For component separation purposes, it is clearly
advantageous to have multiple frequency maps with comparable signal-to-noise
ratio per astrophysical component, as instrumental artefacts are then much
easier to identify. Employing the full set of \Planck\ frequency maps in a
future analysis will improve these results, and provide many more internal
cross-checks. However, we once again stress that the main aim of the current
study is not to present a new state-of-the-art sky model, but rather to demonstrate
the \BP\ algorithm.

\begin{figure*}[t]
\center
\includegraphics[width=.49\linewidth]{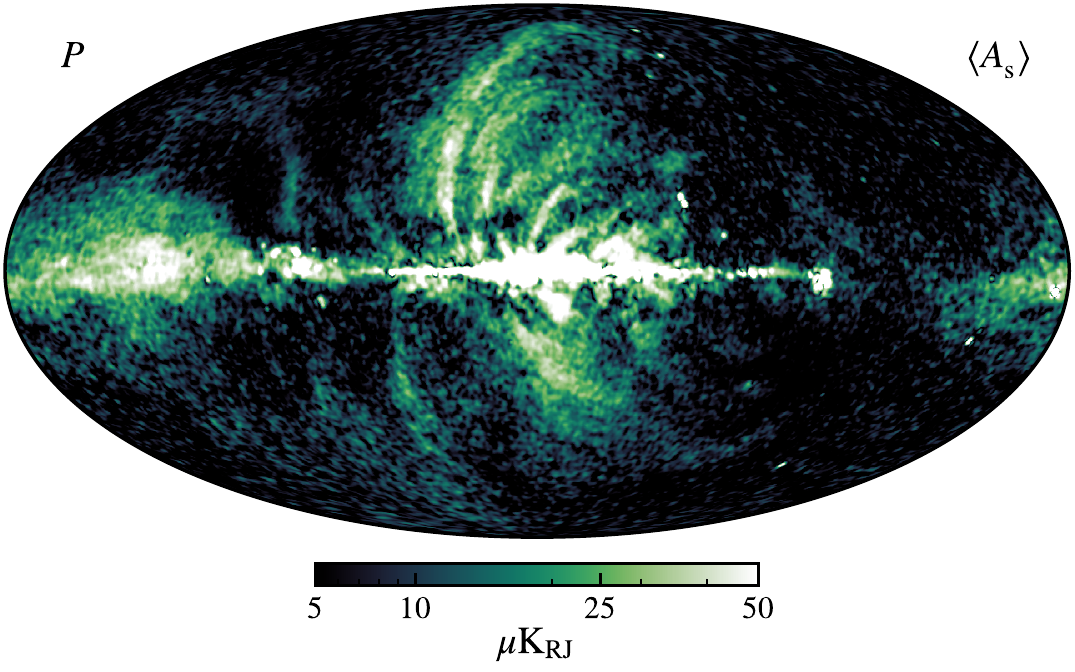}
\includegraphics[width=.49\linewidth]{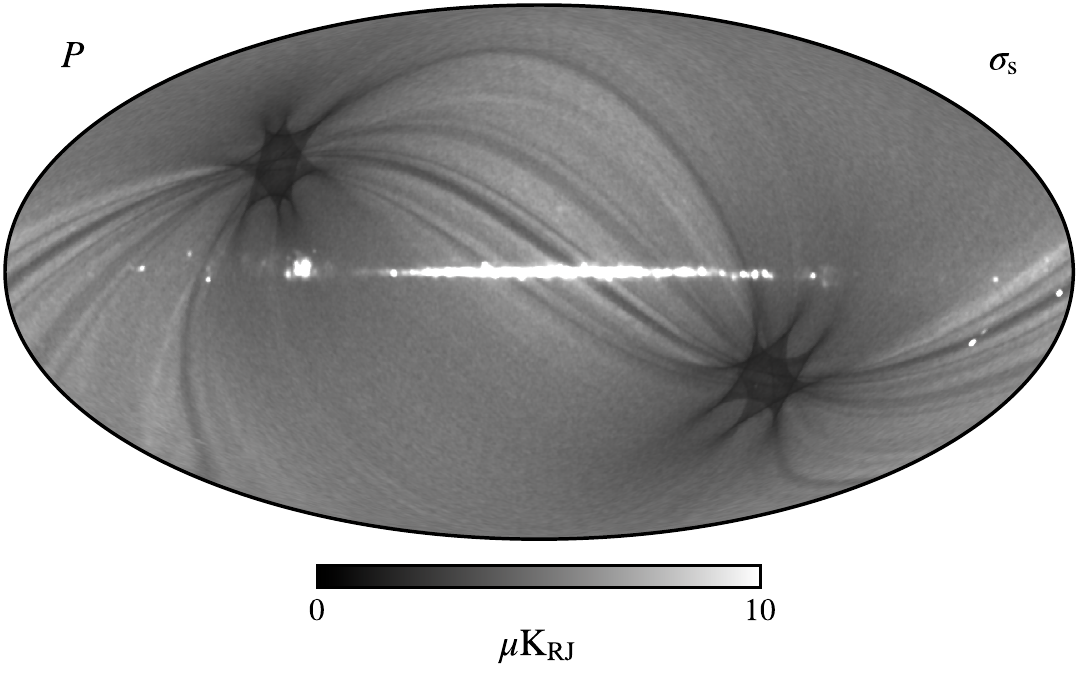}\\
\includegraphics[width=.49\linewidth]{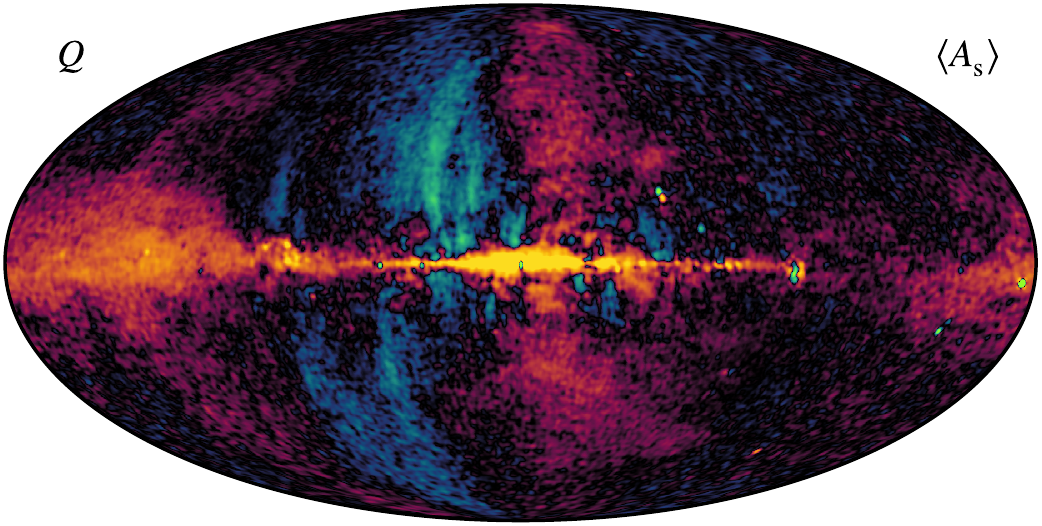}
\includegraphics[width=.49\linewidth]{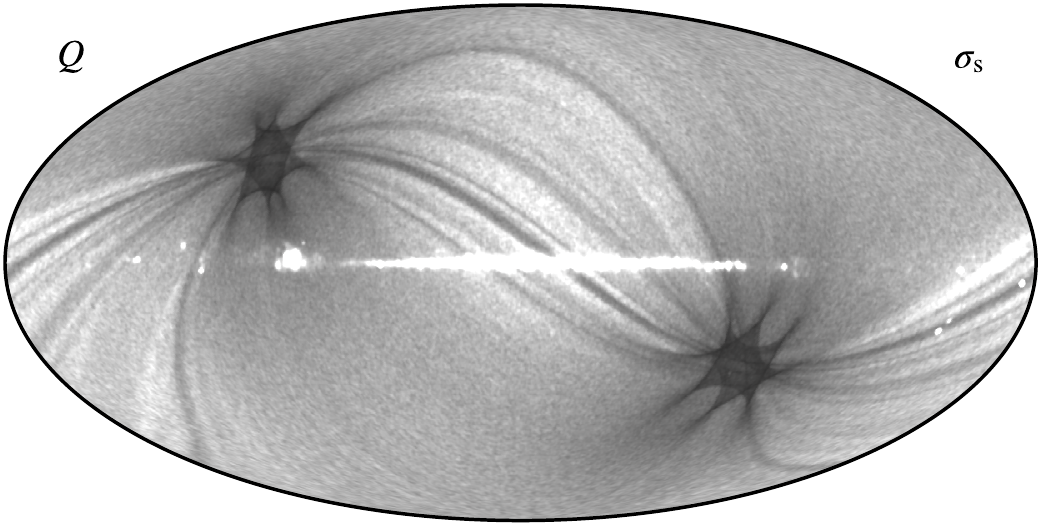}\\
\includegraphics[width=.49\linewidth]{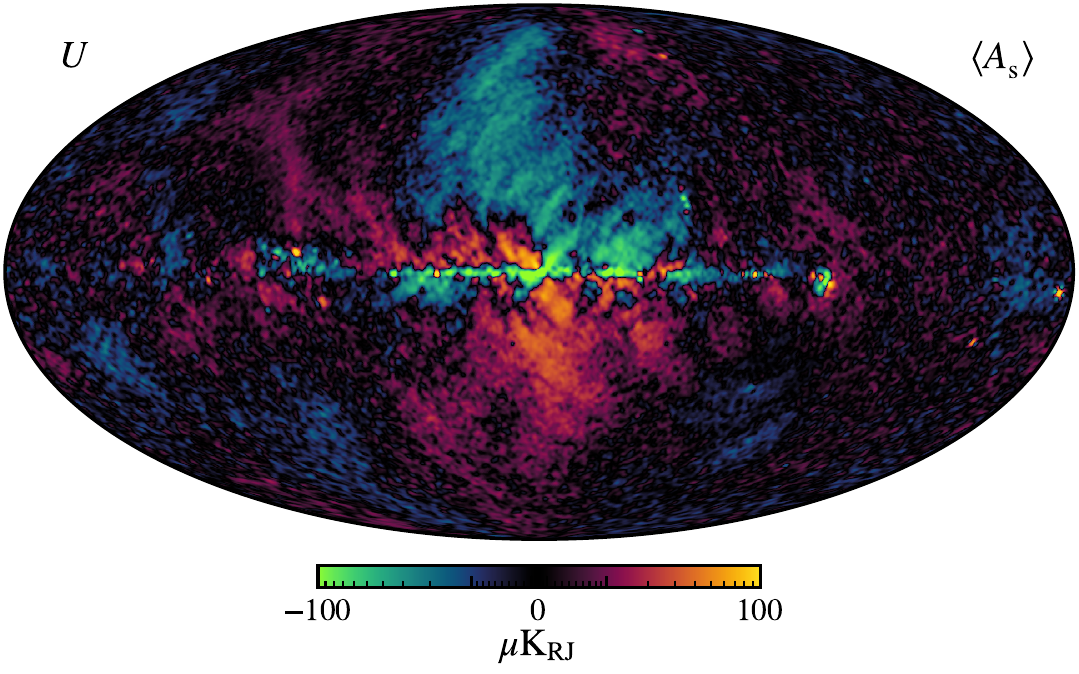}
\includegraphics[width=.49\linewidth]{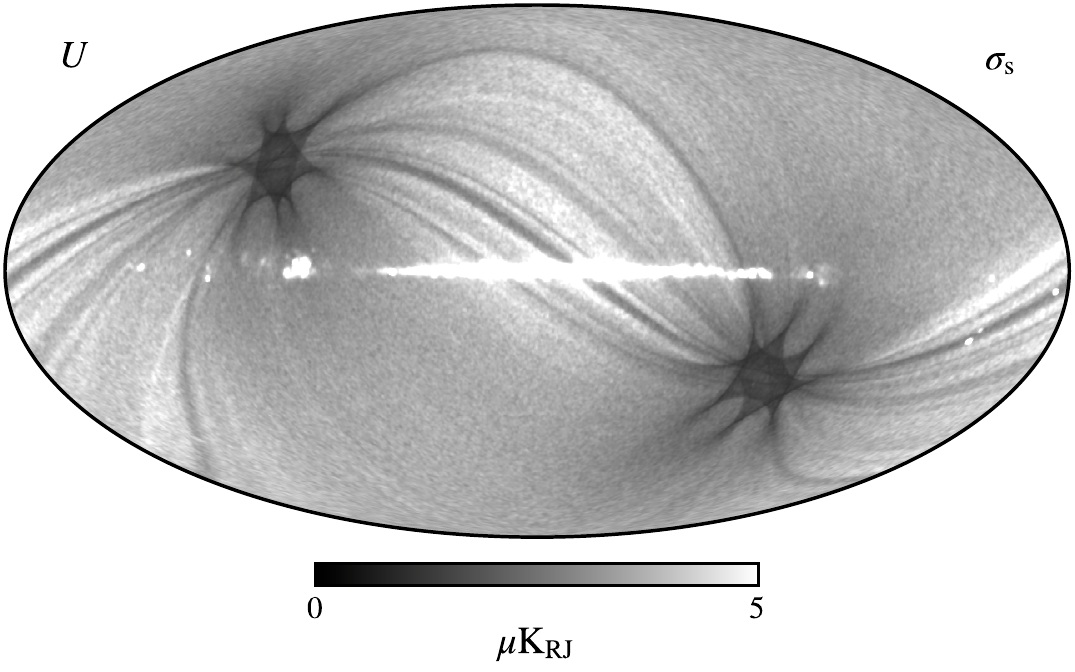}
\caption{Same as Fig.~\ref{fig:Ad}, but for polarized synchrotron
  emission evaluated at 30\,GHz. In this case, the effective angular
  resolution is $1^{\circ}$ FWHM.
}\label{fig:As}
\end{figure*}
\begin{figure*}
\includegraphics[width=0.49\linewidth]{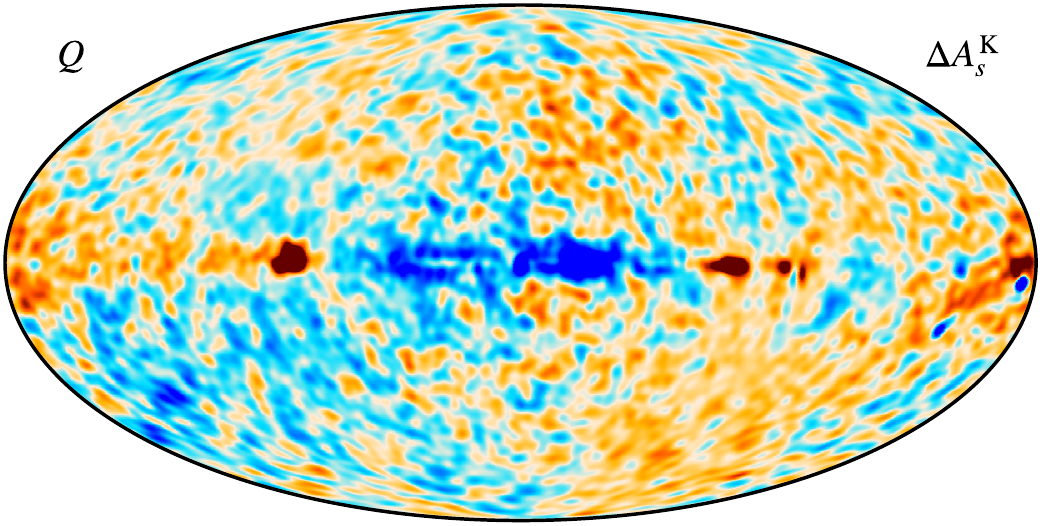}
\includegraphics[width=0.49\linewidth]{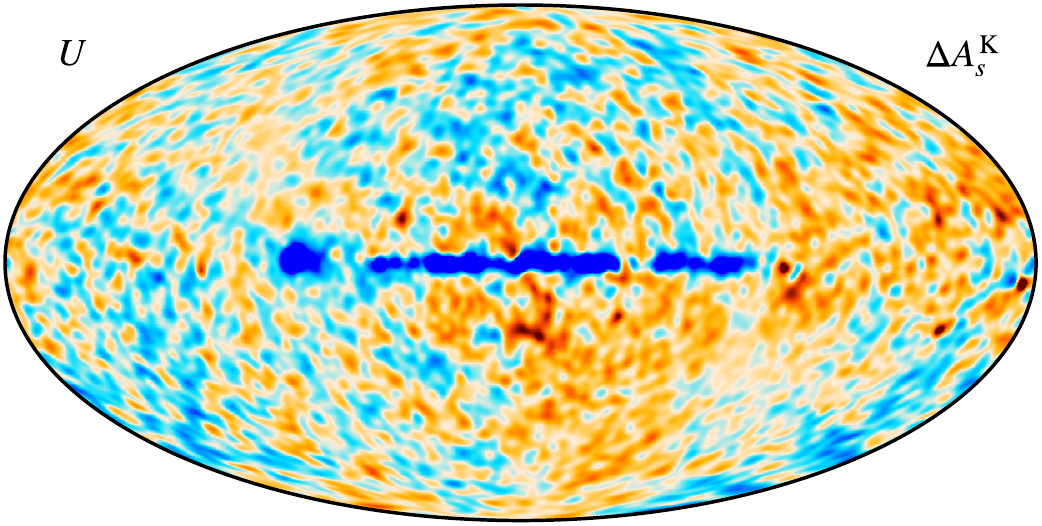}\\
\includegraphics[width=0.49\linewidth]{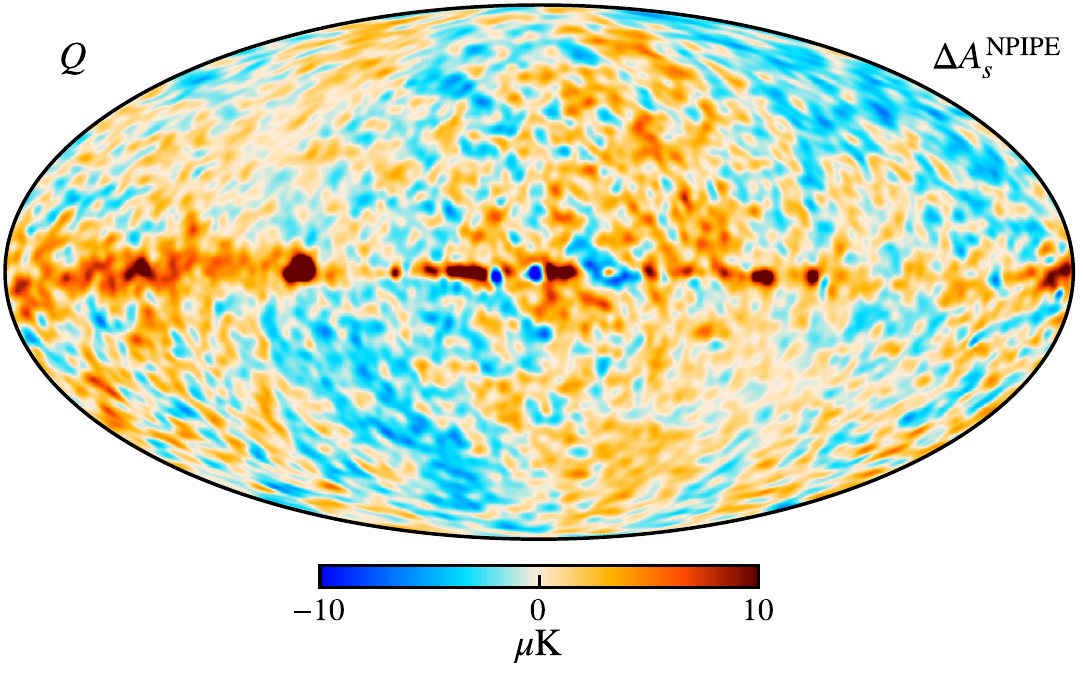}
\includegraphics[width=0.49\linewidth]{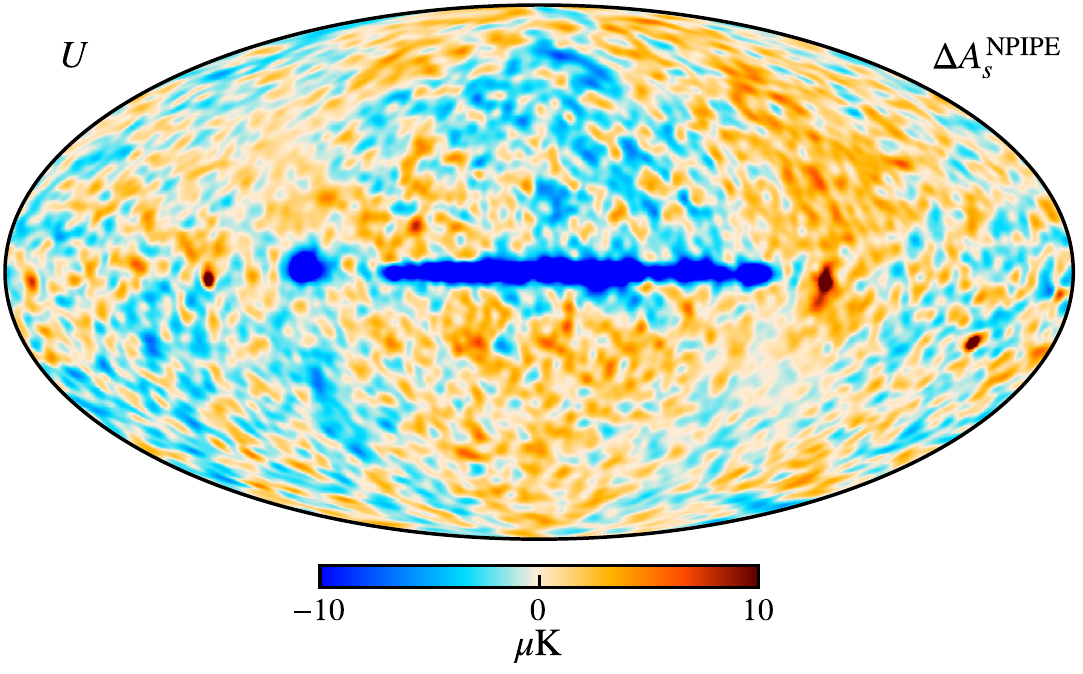}
\caption{(\textit{Top panel:}) Difference between the \BP\ polarized synchrotron
amplitude and the raw \WMAP\ $K$-band map \citep{bennett2012}, the latter being
scaled to 30\,GHz assuming a spectral index of $\beta_s=-3.1$. (\textit{Bottom
panel:}) Similar difference between the \BP\ and \Planck\ DR4 \citep{planck2020-LVII}
synchrotron amplitude maps. Left and right columns show Stokes $Q$ and $U$
parameters respectively, and all maps are smoothed to a common angular
resolution of $3^{\circ}$ FWHM.}\label{fig:Kvalid}\label{fig:sdiff}
\center
\begin{overpic}[width=\linewidth]{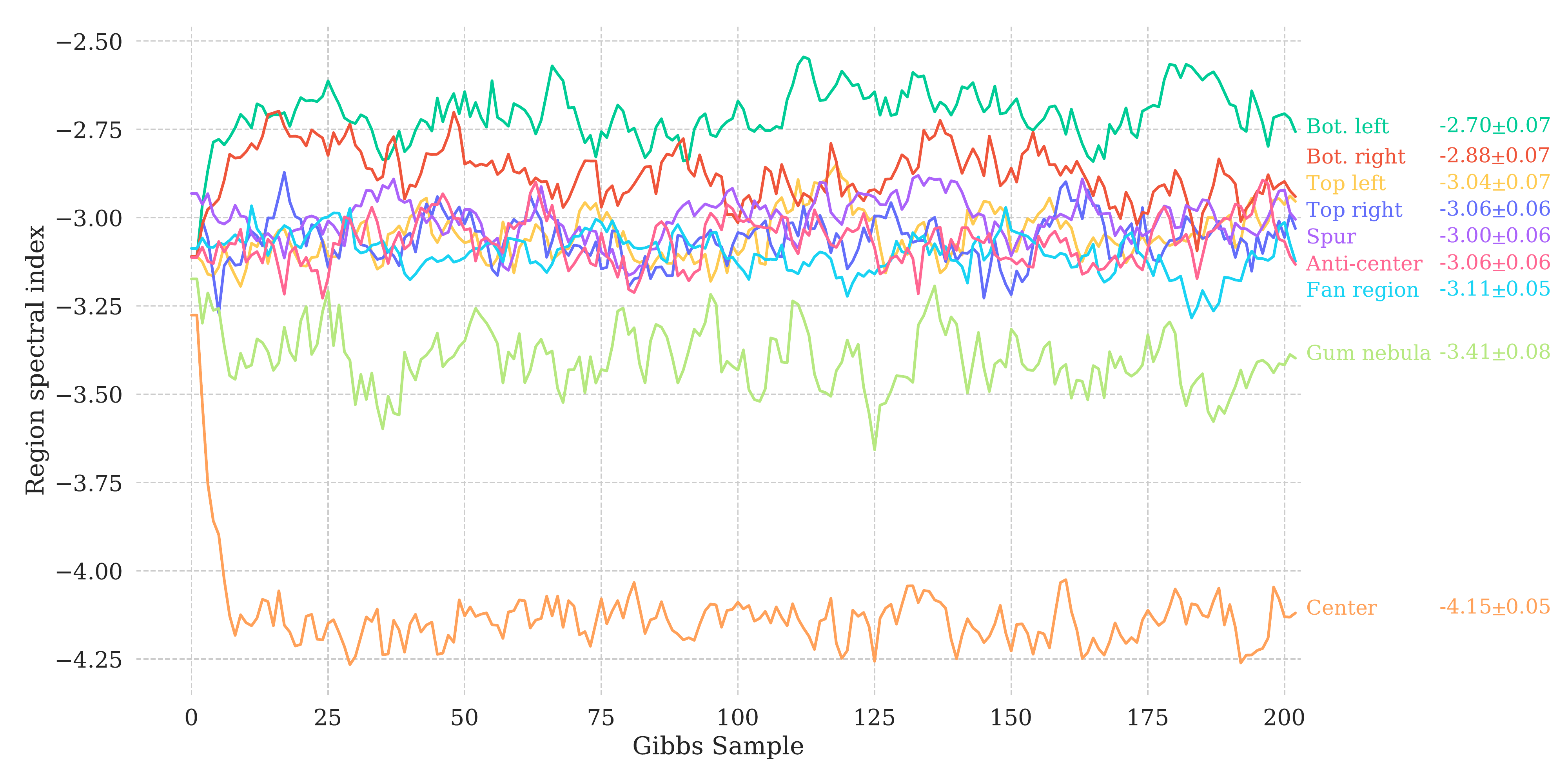}
\put(83,15.5){\includegraphics[scale=0.4]{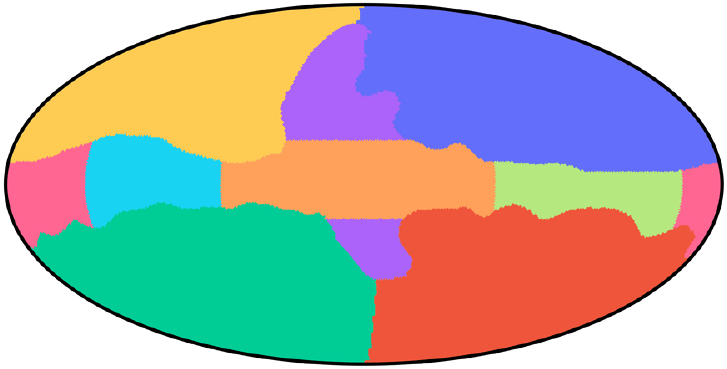}}  
\end{overpic}
\caption{Synchrotron spectral index as a function of Gibbs iteration for an analysis
that includes nine disjoint regions (see inset sky map). The prior is in this case
$\beta_{\mathrm{s}}=-2.8\pm 0.1$, and all regions are constrained using the
full posterior distribution. }\label{fig:synch_trace_t3}
\end{figure*}

Figure~\ref{fig:reswmap} shows the remaining residual maps, i.e.,
\WMAP\ \textit{Ka}, $Q$, and $V$-band. In all three cases, we see
coherent large-scale structures that are clearly morphologically
inconsistent with instrumental noise. These residual maps appear
visually similar to the set of correction templates presented by
\citet{jarosik2007} that account for known transmission imbalance
between the A- and B-sides of the \WMAP\ instrument,\footnote{Note
  that \citealp{jarosik2007} provide two correction templates per
  \WMAP\ differencing assembly, and both $Q$- and $V$-bands are
  therefore associated with four templates each. Only one of these are
  shown in Fig.~\ref{fig:reswmap} for intuition purposes; the other
  templates look qualitatively similar.} as shown in the second and
fourth column of Fig.~\ref{fig:reswmap}.  In our study, we do not
apply any explicit corrections for these templates; hence, they appear
in the frequency residual maps. However, they are accounted for in the
\WMAP\ covariance matrices, and the corresponding modes are therefore
appropriately down-weighed when fitting the astrophysical parameters
with Eqs.~\eqref{eq:ampl_samp_wiener} and
\eqref{eq:beta_posterior}. Transmission imbalance effects will
therefore not bias any astrophysical results, but only result in
larger uncertainties. At the same time, these residual maps clearly
suggest how a future joint time-domain processing of \WMAP\ and
\Planck\ will be able to constrain the \WMAP\ transmission imbalance
parameters with high precision. Once that happens, the corresponding
spatial modes will no longer need to be algebraically projected out,
as is effectively done now, but may rather be used for scientific
inference, on the same footing as any other mode. This work has
already started, and preliminary results are discussed by
\citet{bp17}.

\subsection{Amplitude results}

We now turn our attention to the astrophysical results, and start with
the posterior amplitude maps, $\a$. First, Fig.~\ref{fig:Ad} shows the
thermal dust amplitude, plotted at an angular resolution of
$10\arcmin$ FWHM at $\nside\,$=$\,1024$ for the polarization amplitude
${P=\sqrt{Q^2+U^2}}$, as well as $Q$ and $U$ averaged over an ensemble
of Gibbs samples. The right column shows the corresponding posterior
distribution standard deviation per pixel, with values peaking around
$3\muK_{\mathrm{RJ}}$. As mentioned in the previous subsection, the
thermal dust amplitude is for all practical purposes determined by the
pre-computed HFI 353\,GHz band in the \BP\ processing, and the LFI
bands have little influence. As a result, the thermal dust standard
deviation maps shown in Fig.~\ref{fig:Ad} are essentially given
deterministically by the input 353\,GHz standard deviation
map. These uncertainties are underestimated in the Galactic
plane, where systematic effects must be significant. Overall,
the polarized thermal dust amplitude map is in good
agreement with previous results.

Figure~\ref{fig:As} shows corresponding results for the polarized
synchrotron amplitude map. At high Galactic latitudes, we see that the
standard deviation of this component traces the LFI 30\,GHz scanning
strategy, and is correspondingly largely determined by instrumental
white noise. However, in this case the Galactic plane is in fact
dominated by temperature-to-polarization leakage due to bandpass and
gain uncertainties, resulting in a morphology that matches the 30\,GHz
intensity map.

In Fig.~\ref{fig:sdiff}, we show Stokes parameter maps of the difference between
the \BP\ map of polarized synchrotron amplitude and two independent synchrotron
tracers. The top panel shows differences with respect to the full \WMAP\
$K$-band frequency map at $23\GHz$. To account for its different effective
frequency, we scale the $K$-band map according to a power-law model with
$\beta_{\mathrm s}=-3.1$, or, explicitly, by a factor of 0.38. The two data sets
appear to agree reasonably well, with a certain degree of diffuse large scale
structure. Considering that the effective frequency difference between
$K$-band and 30\,GHz is only $7\GHz$, spatial variations in the
synchrotron spectral index are unlikely to be relevant, as a
difference of $\Delta\beta_{\mathrm s}=0.02$ only translates into
approximately $0.5\muK$ in the relevant regions.

The bottom panel of Fig.~\ref{fig:sdiff} shows similar difference maps with
respect to the polarized synchrotron amplitude map derived from \Planck\ DR4
\citep{planck2020-LVII}. In this case, the high-latitude residuals are dominated
by the \Planck\ scanning strategy, with an overall morphology that closely
matches the LFI gain residual template produced by \citet{planck2016-l02} and
discussed by \citet{bp07} in a \BP\ context. At the same time, no striking
correlations are seen between the $K$-band and \Planck\ DR4 residual maps, and
the overall levels of variation in these two difference maps are comparable. 

\subsection{Spectral index results}

\begin{figure*}[t]
\begin{overpic}[width=\linewidth]{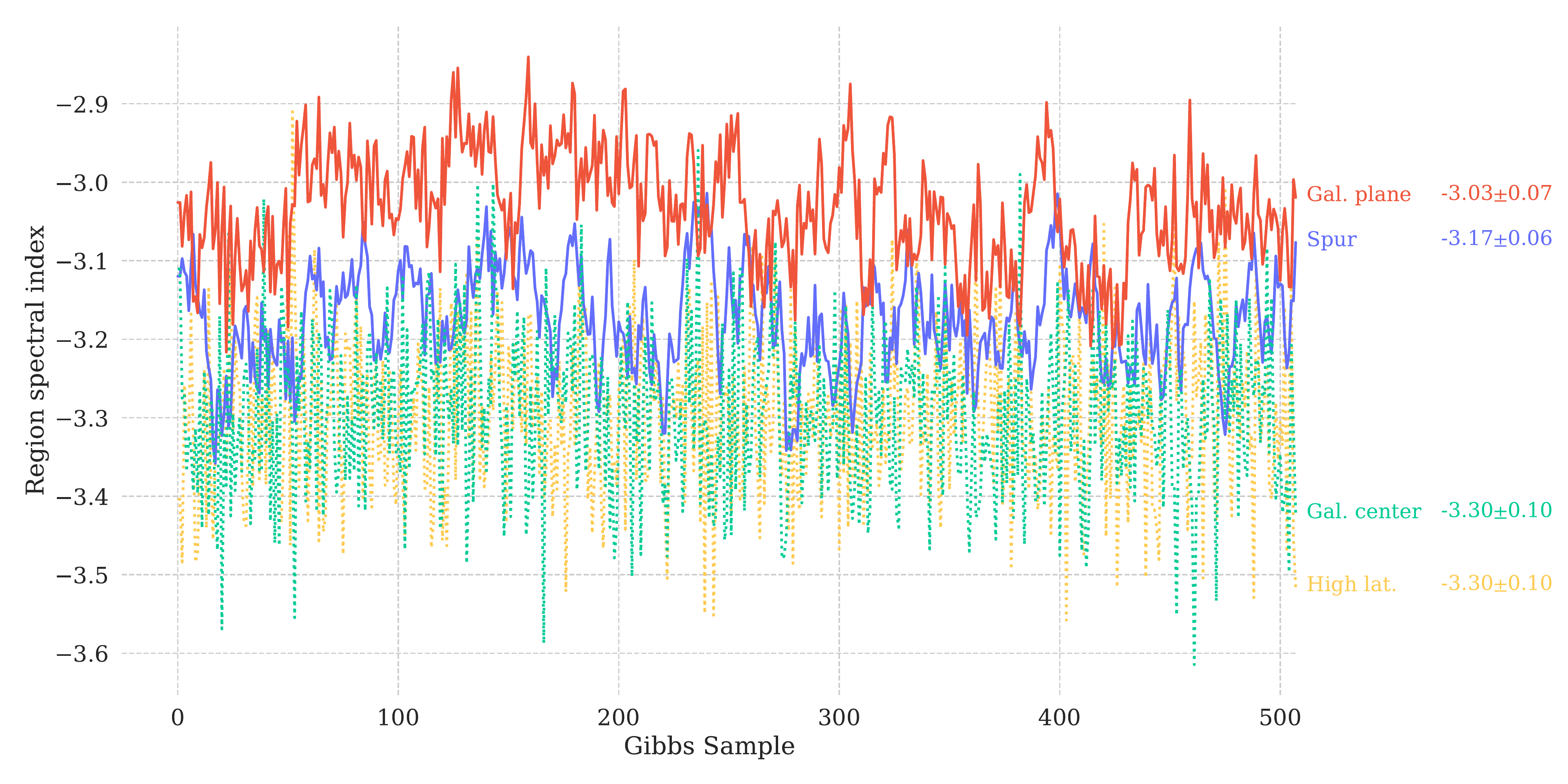}
\put(83,22){\includegraphics[scale=0.4]{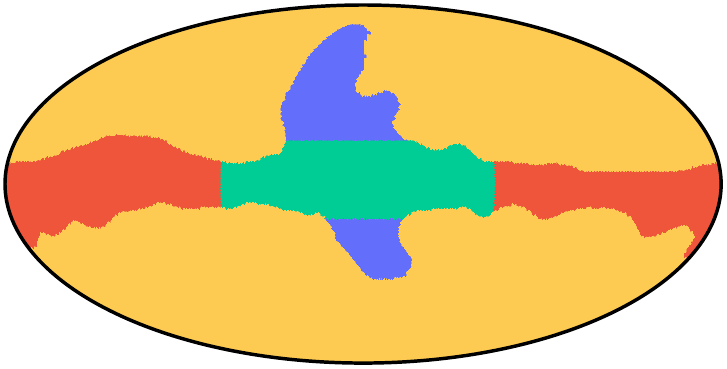}}  
\end{overpic}
\caption{Synchrotron spectral index as a function of Gibbs iteration
  binned using the final \BP\ analysis configuration with four
  disjoint regions and a prior of $\beta_{\mathrm{s}}=-3.3\pm 0.1$. Dotted lines indicate
  regions that are sampled exclusively from the prior distribution,
  while solid lines indicate regions that are sampled from the full
  posterior distribution. }\label{fig:synch_trace_BP8r}
\end{figure*}

\begin{figure*}
\includegraphics[width=0.49\linewidth]{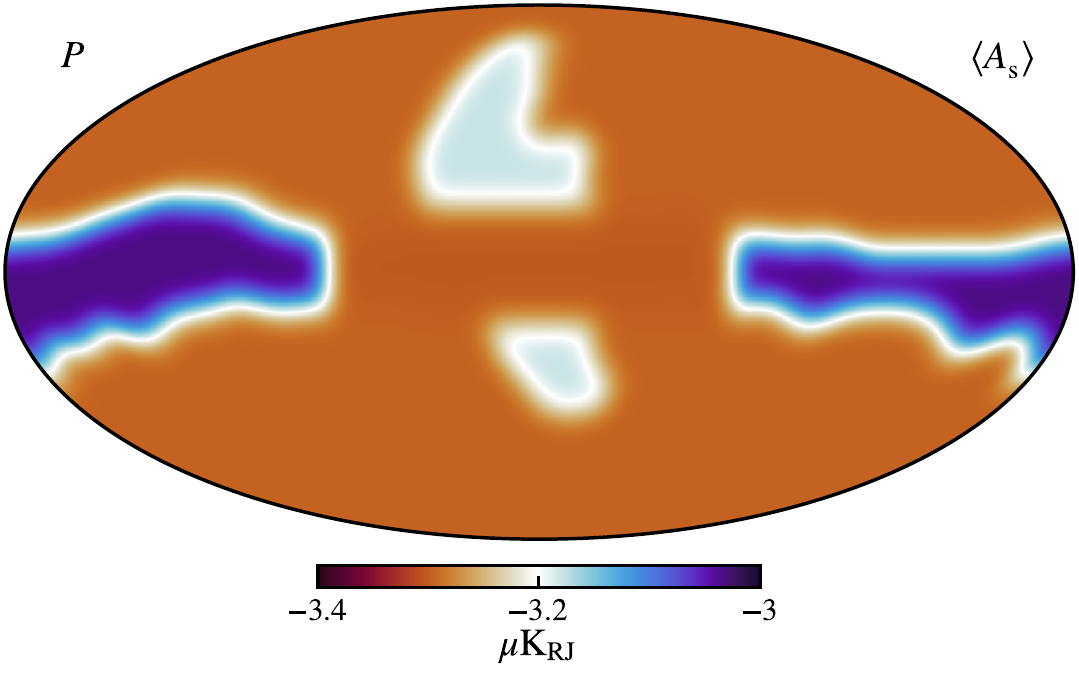}
\includegraphics[width=0.49\linewidth]{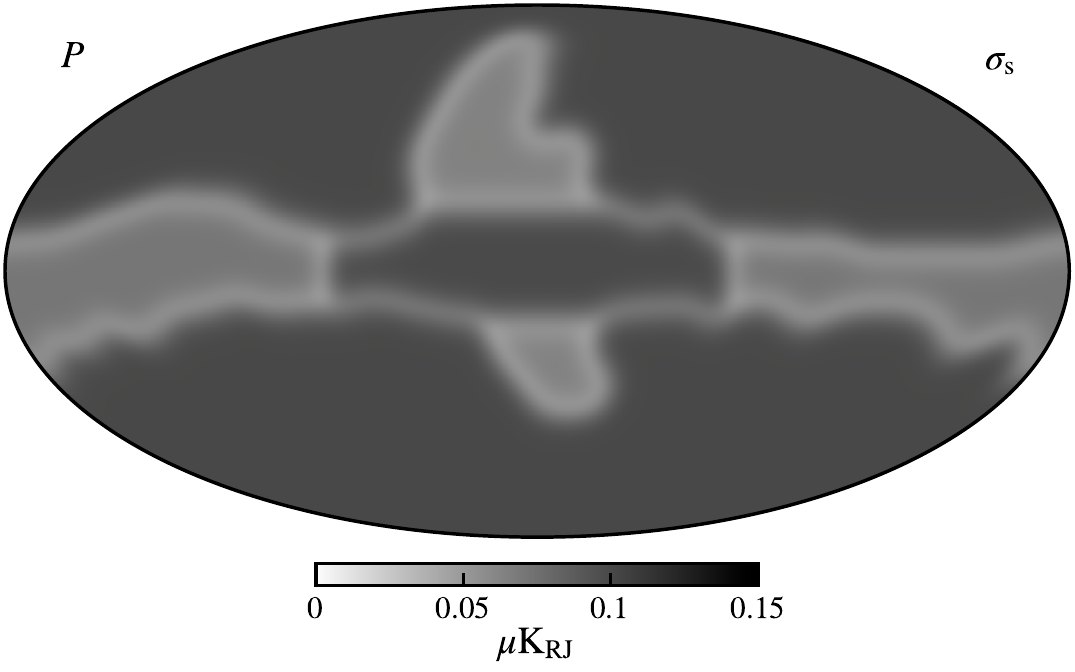}
\caption{Posterior mean (\emph{left panel}) and standard deviation 
  (\emph{right panel}) maps of the spectral
index of polarized synchrotron emission. Note that a prior of
$\beta_{\mathrm{s}}=-3.3\pm0.1$ is applied to all four regions, but only
the Galactic Spur and Galactic Plane regions are constrained
with data through the likelihood; see Sect.~\ref{sec:synch_beta} for further discussion.}\label{fig:synchbeta}
\end{figure*}

\begin{figure}[t]
\center
\includegraphics[width=\linewidth]{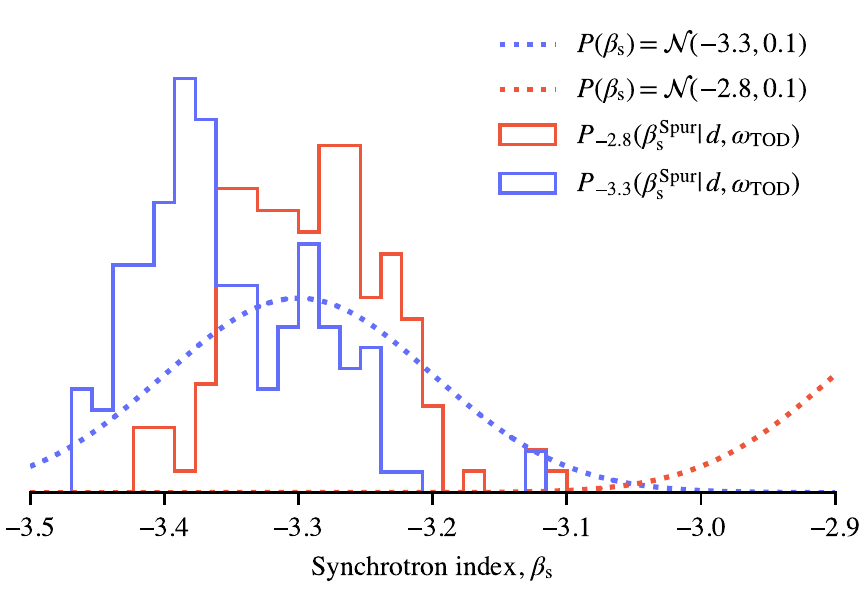}
\caption{Normalized histogram of synchrotron spectral index ($\beta_{\mathrm s}$) for
the Spur region using two different priors. The solid lines show the marginal distribution of spectral index values without TOD sampling using a prior (dotted lines) of $\beta_{s}=-2.8$ (red), and $\beta_{s}=-3.3$ (blue).}
\label{fig:synchbetaspurpriorhist}
\vspace*{3.23mm}
\end{figure} 
  
\begin{figure}[t]
\center
\includegraphics[width=\linewidth]{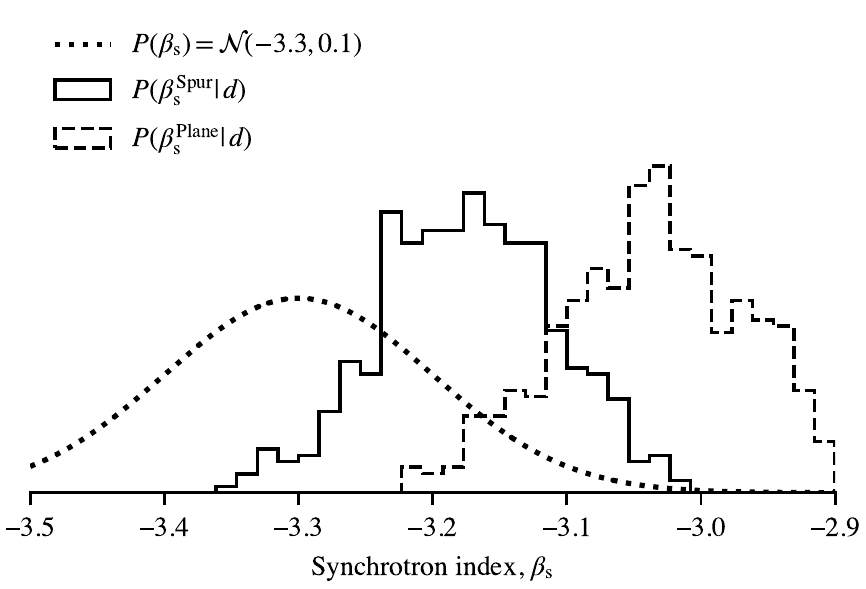}
\caption{Normalized histogram of synchrotron spectral index ($\beta_{\mathrm s}$) for the Spur and Plane regions over the 500 ensemble Gibbs samples, with corresponding prior, $P(\beta_s)$.}\label{fig:synchbetahist}
\end{figure}
 
\subsubsection{Spectral index regions}
\label{sec:regions}

Before presenting the synchrotron spectral index posterior
distribution, we revisit the final choice of sampling regions and
priors by considering a preliminary analysis configuration with nine
disjoint regions rather than four, and a spectral index prior of
$\beta_{\mathrm{s}}=-2.8\pm0.1$, rather than
$\beta_{\mathrm{s}}=-3.3\pm0.1$. The results are summarized in
Fig.~\ref{fig:synch_trace_t3}, which shows $\beta_{\mathrm{s}}$ as a
function of iteration for a Gibbs chain that explored only the
$\{\a,\beta\}$ sub-space.  That is, all TOD-related parameters were
kept fixed in this particular run, in order to highlight the
conditional signal-to-noise ratio of the foreground sector. Each
region is labeled with its corresponding posterior mean and standard
deviation after a burn-in period of 25 Gibbs samples, and accompanied
by a color-coded region map for visualization purposes.

Starting with the most visually striking result, we see that the Galactic Center
region immediately converges to a very low mean value of
$\beta_{\mathrm{s}}=-4.15\pm0.05$, well outside the range of previously reported
values. Of course, we already know that this region is associated with high
$\chi^2$ values (see Fig.~\ref{fig:chisq}), and in particular shows clear evidence of
temperature-to-polarization leakage when compared to \WMAP\ and \Planck\ DR4 (see
Fig.~\ref{fig:sdiff}). Rather than letting these systematic errors potentially
contaminate other parameters through an unphysical synchrotron spectral index
fit, we instead assign the spectral index of this region a physically meaningful
value, as defined by the prior. However, we do not fix it, but rather draw a new
value from the prior in every sample, and thereby marginalize over the prior.
Technically speaking, this is done by omitting the likelihood term in
Eq.~\eqref{eq:beta_posterior} when computing the Metropolis acceptance
probability.

Next, we see that the Gum Nebula region also converges to a notably
low value, and this is most likely related to the same $\chi^2$ excess
that is seen for the Galactic Center. Furthermore, we note that the
Gum Nebula region, as outlined in Fig.~\ref{fig:regions}, exhibits
very little synchrotron signal and is therefore particularly prone to
residual systematic bias when applying weak prior constraints.

Most of the remaining regions fluctuate around values that are at least
nominally consistent with previous analyses reported in the literature. We do
note, however, that the two Southern Hemisphere regions return values that are
high, at $\beta_{\mathrm{s}}=-2.70\pm0.07$ and
$\beta_{\mathrm{s}}=-2.88\pm0.07$, respectively, while most of the remaining
ones lie around $\beta_{\mathrm{s}}\approx-3.1$. Thus, even when measured
conditionally with respect to the TOD parameters, there is slight evidence of a
positive bias of $\Delta\beta_{\mathrm{s}}\approx0.1$ or more with respect to
the Northern Hemisphere. Returning once again to Fig.~\ref{fig:regions}, we see
that all high-latitude regions exhibit very low signal-to-noise ratio, as the
instrumental noise is comparable to, or dominates over, the synchrotron
amplitude in most pixels. These regions are therefore all particularly
susceptible to residual systematic errors and/or prior volume effects
\citep[e.g.,][]{dunkley2009}. Indeed, when sampling jointly over the full range
of time-ordered systematic corrections, we find that these regions converge to
$\beta_{\mathrm{s}}\gtrsim-2.5$. As in the case of the Galactic Center, we do
not take this as evidence for a truly flatter spectral index in these regions,
but rather as an indication that there are low-level residual systematics
present in either \BP, \WMAP, or \Planck\ DR4 353\,GHz (or, possibly, all of them) at
a level that is sufficient to bias the spectral index in the faintest regions. 

We conclude that dividing into nine sky regions is sub-optimal (not to
mention 24, which was the starting point of the analysis; see
Sect.~\ref{sec:synch_model}), as most regions do not have sufficient
signal-to-noise ratio to independently constrain $\beta_{\mathrm{s}}$,
so they are highly susceptible to residual systematic
uncertainties. In particular, small changes in the absolute
calibration can introduce confusing CMB dipole leakage at high
latitudes, while bandpass variations can cause problems near the
Galactic Center. 

We therefore reduce the number of disjoint regions, and
combine the four high Galactic latitude regions into one, and we merge
all regions along the galactic plane, except for the Galactic
center. Even in this minimal configuration, the high Galactic latitude
region is not well constrained, and we therefore sample this from the
prior alone, as for the Galactic Center. This leaves only two regions
(the Galactic Spur and the Galactic Plane) to be sampled
properly with the full posterior distribution, and, fortunately, both
of these appear to be both signal-dominated and stable with respect to
instrumental parameter variations.

\subsubsection{Synchrotron spectral index results}
\label{sec:synch_beta}

We are now finally ready to present the main results of this paper,
namely constraints on the synchrotron spectral index. Starting with
the individual Gibbs samples, Fig.~\ref{fig:synch_trace_BP8r} shows
traceplots for each of the four regions. The first half shows the
first Gibbs chain, and the second half shows the second Gibbs chain,
both after discarding 10 samples for burn-in. Overall, we see that the
correlation length is modest, as in 30--50 samples. The chains mix
well, and appear at least visually to be statistically stationary; it
is not easy to identify the point at which the two chains are joined,
which typically is the case if there are long-term drifts. For
completeness, Fig.~\ref{fig:synchbeta} shows posterior mean and
standard deviation sky maps evaluated from these chains.

We also see that the Spur and Galactic Plane regions, which are the
only two regions for which $\beta_{\mathrm{s}}$ is actually fitted,
have shallower mean spectral indices than the prior of
$\beta_{\mathrm{s}}=-3.3$. This may seem somewhat paradoxical, since
it is then natural to ask why the prior was not set to
$\beta=-3.1$. As discussed earlier, the reasons are two-fold. Firstly,
the High Latitude region actually does appear to prefer a steeper
spectral index, as indicated by the \Planck\ likelihood analysis
\citep{planck2016-l05}, AME constraints \citep{bp15}, and our own
preliminary studies. At the same time, this region is also both the
most important region for CMB purposes, and it is prior-dominated. It
is therefore particularly important that the prior works well for this
region. Secondly, the prior only has a very mild effect on the two
signal-dominated regions anyway, precisely because of their higher
statistical weight. This is explicitly demonstrated in
Fig.~\ref{fig:synchbetaspurpriorhist}, which compares
conditional distributions for the Spur region with two different
priors centered on $\beta=-3.3$ (blue) and $-2.8$ (red),
respectively. (In this case, the TOD parameters are kept fixed, and
the $P(\a,\beta|\d, \omega_{\mathrm{TOD}})$ distribution is explored
for one arbitrarily chosen realization of $\omega_{\mathrm{TOD}}$.)
Here we see that shifting the prior mean by as much as
$\Delta_p\beta=0.5$ only affects the posterior by
$\Delta\beta\lesssim0.1$. For a more realistic possible prior shift of
$\Delta_p\beta=0.2$, the final posterior shifts will be
$\Delta\beta\lesssim0.03$, which is small compared to the overall
variations seen in Fig.~\ref{fig:synch_trace_BP8r}. In short, the Spur
and Galactic plane regions are signal-dominated, and the prior is of
limited importance.

\begin{figure}[t]
  \center
  \vspace*{0.25mm}
  \includegraphics[width=\linewidth]{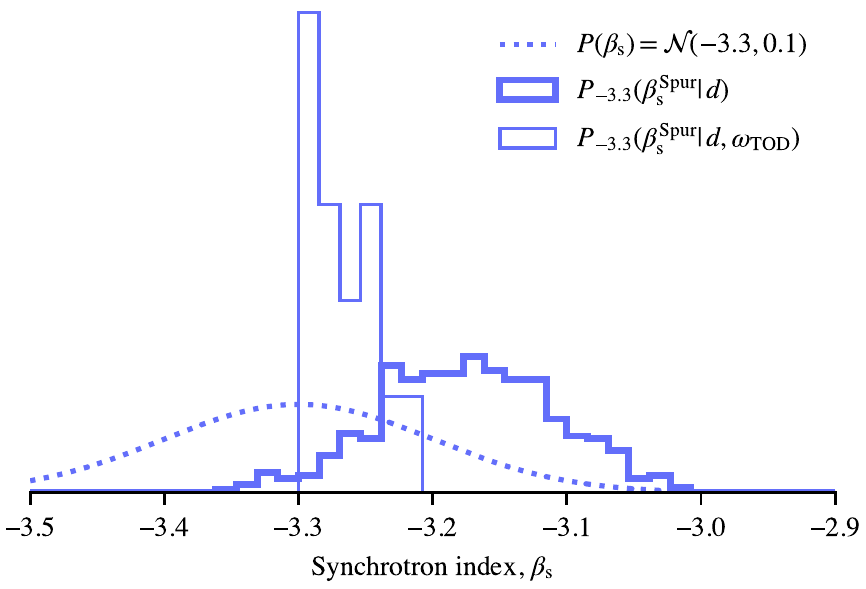}
  \caption{Normalized histogram of synchrotron spectral index
    ($\beta_{\mathrm s}$) for the Spur region using a prior of
    $\beta_{s}=-3.3$. The bold line shows the full
    \BP\ posterior distribution including TOD sampling, and the thin
    line shows the corresponding posterior distribution when conditioning on TOD parameters.}
  \label{fig:synchbetaspurmarghist}
  \end{figure} 

Figure~\ref{fig:synchbetahist} compares the posterior distributions for these
two regions. For these, we find posterior mean and standard deviations of
${\beta_{\mathrm s}^{\mathrm{Spur}}=-3.17\pm0.06}$ and ${\beta_{\mathrm
s}^{\mathrm{Plane}}=-3.03\pm0.07}$. Both of these values are consistent with
earlier constraints in the literature, that suggests a steepening of the
spectral index from low to high latitudes \citep[e.g.,][]{fuskeland:2019, krachmalnicoff2018}. Similarly,
\citet{dunkley2009} reports a variation of $\Delta\beta_{\mathrm s}=0.08$
between low and high Galactic latitudes using the \WMAP\ data, while we find a
variation of $\Delta\beta_{\mathrm s}=0.14$ between the Galactic plane and the
Spur using both \WMAP\ and LFI data. The steepening is however only
statistically significant at the $2\,\sigma$ level as determined in the current analysis. 

Next, to illustrate the importance of marginalization over TOD
parameters, Fig.~\ref{fig:synchbetaspurmarghist} compares the full
marginal posterior distribution (thick blue histogram) with a similar
posterior distribution that fixes the TOD parameters at one arbitrary
Gibbs sample (thin blue histogram). Thus, the former marginalizes over
the full \BP\ data model, while the latter only marginalizes over
foreground parameters and white noise. The relative widths of the two
distributions clearly demonstrates the importance of accounting
uncertainties in the full parameter sets, and correspondingly also the
advantage of joint global parameter estimation.

As an additional validation of these results, we replace the three
\BP-processed LFI frequency map samples with the corresponding
preprocessed \Planck\ DR4 maps \citep{planck2020-LVII}. In this case,
we find spectral indices of
${\beta_{\mathrm{s}}^{\mathrm{Spur}}=-3.20\pm0.06}$ and
${\beta_{\mathrm{s}}^{\mathrm{Plane}}=-3.06\pm0.06}$, respectively,
which are individually statistically consistent with the \BP\ results
at the $0.5\,\sigma$ level. 

\subsubsection{Thermal dust spectral index}

\begin{figure}[t]
\center
\includegraphics[width=0.95\linewidth]{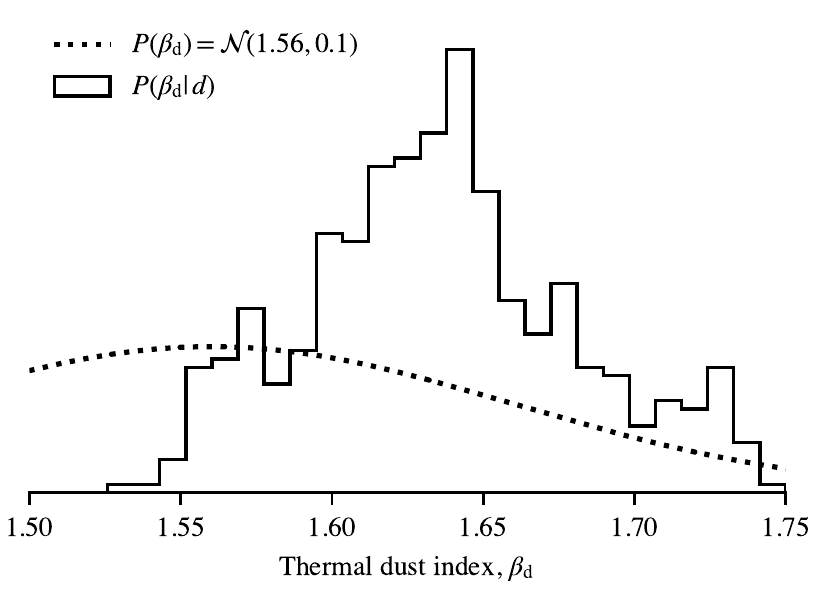}
\caption{Normalized histogram of thermal dust spectral index ($\beta_{\mathrm d}$) over the 500 ensemble Gibbs samples, with corresponding prior.}\label{fig:dustbeta}
\end{figure}
 
For polarized thermal dust emission, we fit only one power-law index,
$\beta_{\mathrm{d}}$, across the full sky, while fixing the dust
temperature on the latest \Planck\ estimate \citep{planck2020-LVII}. This is
exclusively due to a limited signal-to-noise ratio, and not a
statement regarding the complexity of the true sky. In this case, we
adopt a prior of $\beta_{\mathrm{d}}=1.56\pm0.10$, motivated by the
most recent \Planck\ HFI results \citep{planck2016-l04,planck2020-LVII}. 

The resulting posterior distribution is shown in
Fig.~\ref{fig:dustbeta}, which may be reasonably approximated as a
Gaussian with $\beta_{\mathrm d}=1.62\pm 0.04$. This mean value is thus slightly
steeper than expected based on HFI, with a statistical significance of
about $1.5\,\sigma$. Furthermore, the uncertainty is significantly
smaller than the prior width, which suggests that the result is indeed
data-driven, even when marginalizing over the full \BP\ instrument
model.

While we caution against over-interpreting the significance of this
result, we do note that a possible spectral steepening in the thermal
dust SED around 100\,GHz would have dramatic consequences for future
high-sensitivity $B$-mode experiments. Thus, understanding whether
this result is due to a statistical fluke, or instrumental modeling
errors (for instance, because of an overly simplistic bandpass correction
model), or actual astrophysics is an important goal for future
analysis. Including \Planck\ HFI frequencies between 100 and 217\,GHz
in a future analysis will clearly be informative in this respect.

\section{Summary and conclusions}  
\label{sec:conclusions}

The two main goals of this paper are to introduce a Bayesian sampling
algorithm for polarized CMB foreground models as embedded within the
end-to-end \BP\ framework, and to present the first results from this
pipeline as applied to the \Planck\ LFI data set. This is the first
time a joint global parametric model that accounts for both
instrumental and astrophysical parameters has been fit to the
\Planck\ LFI data within a single joint posterior distribution,
allowing for seamless end-to-end error propagation. This is also the
first time a joint analysis of the \Planck\ LFI and \WMAP\ data
has resulted in physically meaningful spectral indices for both polarized
synchrotron and thermal dust emission. This analysis thus paves the
way for future analyses that should integrate and analyze more data
sets.

Indeed, we stress that the current analysis configuration has been
specifically designed to demonstrate the properties and performance of
the algorithm itself, not to derive a new best-fit sky
model. Specifically, critically important data sets, such as
\WMAP\ $K$-band and \Planck\ HFI, have been intentionally omitted,
precisely because of their high signal-to-noise ratios; if they had
been included, the results would have been dominated by \WMAP\ and HFI
methodology. Including these data sets, and other important ones like
C-BASS \citep{jew2019} or QUIJOTE \citep{QUIJOTE_I_2015} will be done
in future work, either by members within the current \BP\ team or by
external researchers, and either by starting from time-ordered or from
pre-pixelized sky maps. In many respects, the current analysis
configuration may very possibly represent one of the most difficult
challenges that the \BP\ pipeline will face, since it is
the least constrained; future analyses will always have access to more
data, and the resulting sky models will therefore be less degenerate.

With that important caveat in mind, particularly notable highlights
from the current analysis include the following:
\begin{enumerate}
\item We constrain the spectral index of polarized synchrotron emission,
  $\beta_{\mathrm{s}}$, in two large and disjoint regions of the sky, covering
  the Galactic Spur and the Galactic Plane, with best-fit values $\beta_{\mathrm
  s}^{\mathrm{Spur}}=-3.17\pm 0.06$ and $\beta_{\mathrm
  s}^{\mathrm{Plane}}=-3.03\pm 0.07$, respectively. These results are
  statistically consistent with previous \WMAP-only results \citep{dunkley2009}.
  The current analysis finds some evidence for spatial variation between spur
  and plane in $\beta_{\mathrm{s}}$, but only statistically significant at the
  $2\,\sigma$ level. At the same time, we note that the high Galactic latitude
  region has a too low signal-to-noise ratio to support any robust conclusions
  regarding $\beta_{\mathrm{s}}$, while the Galactic Center region exhibits too
  strong residual systematic effects.
\item We constrain the spectral index of thermal dust emission between
  30 and 70\,GHz to $\beta_{\mathrm d}=1.62\pm 0.04$, which is somewhat steeper
  than that previously reported by \Planck\ HFI \citep{planck2016-l04,
    planck2020-LVII} of $\beta_{\mathrm{d}}\approx 1.56$, but still
  statistically consistent. 
\item Through joint analysis of the \Planck\ and \WMAP\ data, we have
  been able to isolate and highlight the effect of transmission
  imbalance in the \WMAP\ observations. This strongly suggests that a
  future joint analysis of \Planck\ and \WMAP\ data in the time-domain
  will be able to constrain the \WMAP\ transmission imbalance
  parameters to high accuracy. This work has already started, as
  discussed by \citet{bp17}.
\end{enumerate}

\begin{figure}[t]
  \center
  \includegraphics[width=\linewidth]{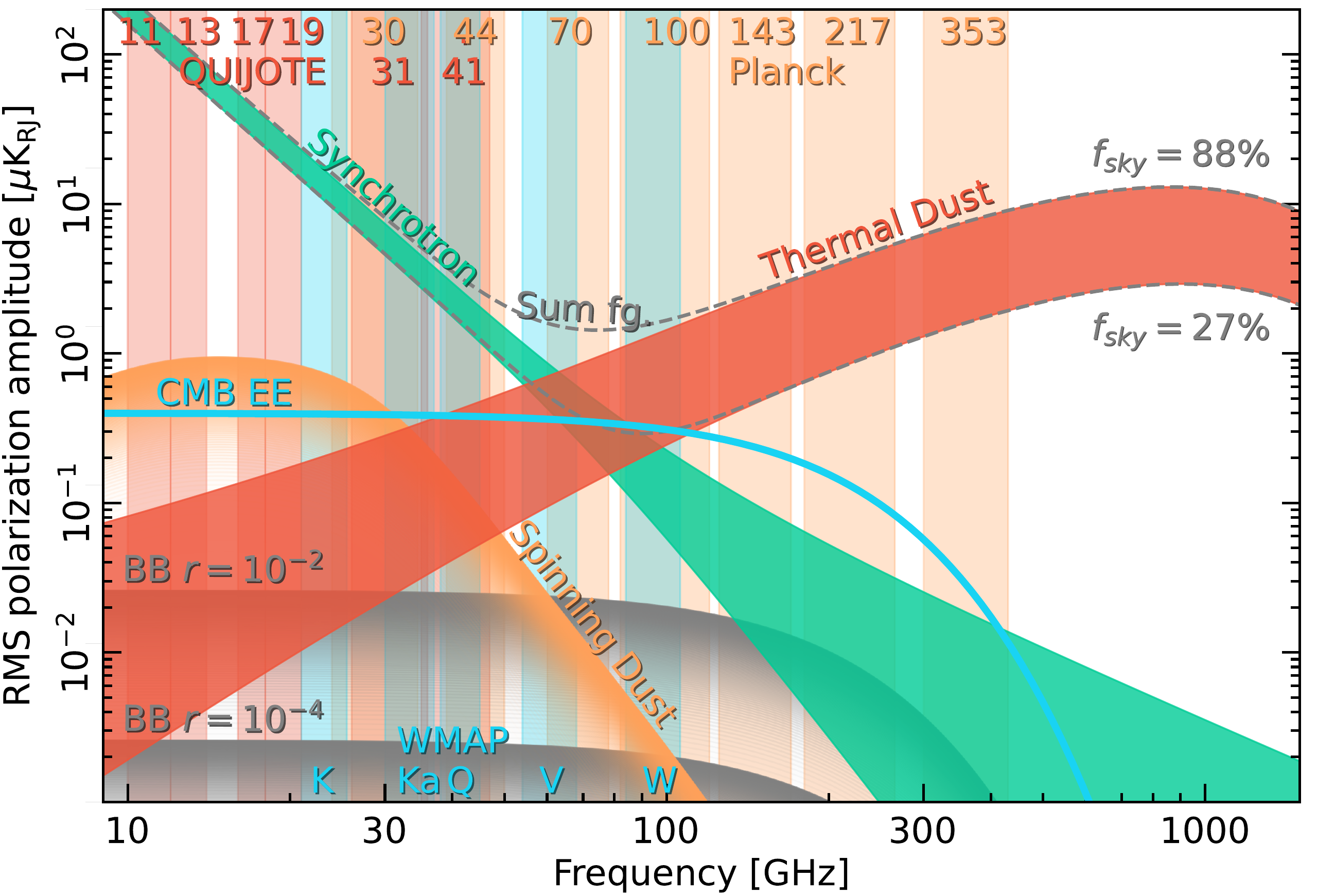}
  \caption{Polarization amplitude root-mean-square (RMS) as a function of
    frequency and astrophysical component and polarized emission. Each component
    map is smoothed to a common angular resolution of $1^{\circ}$ FWHM, and the
    lower and upper edges of  each band are defined by the RMS outside masks
    covering $27\,\%$ and $88\,\%$ of the sky, respectively. An artificial
    broadening has been applied to the low-amplitude tails of the foreground
    spectra to visually indicate uncertainties in low signal-to-noise regions
    for the relevant components. The $EE$ CMB spectrum is generated from ideal
    CMB simulations based on the best-fit \Planck\ $\Lambda$CDM model, while the
    $BB$ limits are derived with tensor-to-scalar ratios of $r=10^{-2}$ and
    $10^{-4}$, respectively. Additionally, we visualize an upper limit on the
    polarization fraction of spinning dust of 1\,\% as has been reported in
    literature \citep{QUIJOTE_II_2016, macellari2011, bp15}.}
  \label{fig:skymodel}
  \end{figure} 
Figure~\ref{fig:skymodel} provides an overview of the main polarized microwave
components in the frequency range from 10 to 1000\,GHz, as described by the
posterior \BP\ results and our assumed sky model with the addition of spinning
dust with a polarization fraction of 1\,\%  \citep{QUIJOTE_II_2016,bp15}. Here, each
component is represented in terms of the standard deviation of the polarization
amplitude evaluated over 88\,\% (top edge of each band) and 27\,\% (bottom edge
of each band) of the sky, with all \WMAP\ and \Planck\ frequency bands marked as
vertical columns. To illustrate the importance of detailed foreground modeling
and error propagation, the predicted levels of CMB $BB$ power for
tensor-to-scalar ratios of $r=10^{-2}$ and $r=10^{-4}$ are marked by diffuse
gray regions. In order to achieve a significant measurement of this signal,
exquisite control over both foreground contamination and systematic effects and
their interplay will be quintessential. We believe that the analysis framework
presented in this paper, and in a suite of companion papers, can play an
important role in this work, by providing a common and statistically
well-defined analysis platform for past, current and future CMB experiments.

\begin{acknowledgements}
  We thank Prof.\ Pedro Ferreira and Dr.\ Charles Lawrence for useful suggestions, comments and 
  discussions. We also thank the entire \Planck\ and \WMAP\ teams for
  invaluable support and discussions, and for their dedicated efforts
  through several decades without which this work would not be
  possible. The current work has received funding from the European
  Union’s Horizon 2020 research and innovation programme under grant
  agreement numbers 776282 (COMPET-4; \BP), 772253 (ERC;
  \textsc{bits2cosmology}), and 819478 (ERC; \textsc{Cosmoglobe}). In
  addition, the collaboration acknowledges support from ESA; ASI and
  INAF (Italy); NASA and DoE (USA); Tekes, Academy of Finland (grant
   no.\ 295113), CSC, and Magnus Ehrnrooth foundation (Finland); RCN
  (Norway; grant nos.\ 263011, 274990); and PRACE (EU).
\end{acknowledgements}

\bibliographystyle{aa}

\bibliography{Planck_bib,BP_bibliography}

\end{document}